\documentclass[prx,superscriptaddress,longbibliography,twocolumn]{revtex4}
\usepackage{appendix}
\usepackage{graphicx}
\usepackage{bm}
\usepackage{bbm}
\usepackage{color}
\usepackage{bm}
\usepackage{amsmath}
\usepackage{siunitx}
\usepackage{amssymb}
\usepackage{mathrsfs}
\usepackage[caption=false]{subfig}
\usepackage{physics}
\usepackage{hyperref}
\usepackage[normalem]{ulem} 

\usepackage{url}
\usepackage{tikz}
\usepackage{pgfplots}

\usepackage[utf8]{inputenc}

\newcommand{\mcU}{\mathcal{U}}

\newcommand{\R}{\mathbb{R}}

\begin{document}

\title{Magnetic skyrmion crystal at a topological insulator surface}
\author{Stefan Divic}
\email{stefan\_divic@berkeley.edu}
\affiliation{Department of Physics, University of California, Berkeley, California 94720, USA}
\author{Henry Ling}
\affiliation{208-5800 Cooney Road, Richmond, British Columbia V6X3A8, Canada}
\author{T. Pereg-Barnea}
\email{tamipb@physics.mcgill.ca}
\affiliation{Department of Physics and the Centre for the Physics of Materials,
McGill University, Montr\'eal, Qu\'ebec H3A 2T8, Canada}
\author{Arun Paramekanti}
\email{arun.paramekanti@utoronto.ca}
\affiliation{Department of Physics, University of Toronto, Toronto, Ontario M5S 1A7, Canada}
\date{\today}

\begin{abstract}
We consider a magnetic skyrmion crystal formed at the surface of a topological insulator. Incorporating the exchange interaction between the helical Dirac surface states and the periodic N\'eel or Bloch skyrmion texture, we obtain the resulting electronic band structure and discuss the constraints that symmetries impose on the energies and Berry curvature. We find substantive qualitative differences between the N\'eel and Bloch cases, with the latter generically permitting a multiband low energy tight-binding representation whose parameters are tightly constrained by symmetries. We explicitly compute the associated Wannier orbitals, which resemble the ringlike chiral bound states of helical Dirac fermions coupled to a single skyrmion in a ferromagnetic background. We construct a two-band tight-binding model with real nearest-neighbor hoppings which captures the salient topological features of the low-energy bands. Our results are relevant to magnetic topological insulators (TIs), as well as to TI-magnetic thin film heterostructures, in which skyrmion crystals may be stabilized.
\end{abstract}

\maketitle

\section{Introduction}
Massless Dirac fermions emerge in condensed matter as low energy excitations of systems whose Fermi levels sit close to a band crossing. Notable examples of this phenomenon in dimensions $d > 1$ include graphene, Weyl/Dirac semimetals, and the surface states of strong topological insulators (TIs)  \cite{HasanKaneRMP,QiZhangRMP,WeylRMP,PhysRevLett.98.106803, PhysRevB.75.121306, PhysRevB.79.195321, PhysRevB.78.195424}.
In certain cases, such band touchings may be protected by lattice or time-reversal symmetries, so that breaking these symmetries induces a Dirac mass gap, leading to physically observable consequences \cite{PhysRev.140.A401, PhysRevLett.61.2015, PhysRevB.75.033408, PhysRevB.74.235111, PhysRevLett.95.226801}. For instance, inducing a mass gap in graphene by breaking inversion symmetry leads to a valley Hall effect \cite{ValleyHall}, while breaking time-reversal symmetry for a TI surface Dirac cone via a perpendicular Zeeman field leads to a gapped half-integer quantum Hall state \cite{QAHreview} with $\sigma_{xy}=e^2/2h$. Such symmetry breaking may be induced by proximity coupling with a symmetry-broken substrate, or by spontaneous ordering of magnetic moments. Domain walls of such broken symmetries, where the Dirac mass changes sign, act as channels which support chiral edge modes \cite{JackiwRebbi,grapheneCDW,TIdw}.

Going beyond the impact of uniform symmetry breaking orders and isolated domain walls, it is interesting to consider the effect of periodically modulated potentials on massless Dirac fermions. Such modulations have been extensively studied in the context of superlattices in graphene \cite{grapheneSL1,grapheneSL2,grapheneSL3,grapheneSL4,grapheneSL5,grapheneSLexpt} and bilayer graphene \cite{BLGSL1,BLGSL2}, where they have been shown to produce emergent Dirac fermion excitations.
The superlattice reconstruction of low-energy bands has also recently come to the fore in studies of twisted bilayer graphene \cite{Bistritzer_TBG, PhysRevB.86.155449, PhysRevB.98.085435, PhysRevLett.109.126801, PhysRevB.81.161405, PhysRevLett.99.256802, doi:10.1021/nl902948m, PhysRevB.81.165105, PhysRevB.82.121407, PhysRevLett.106.126802, PhysRevLett.109.186807, 10.1038/nphys1463, PhysRevLett.117.116804, LedwithPRR, PhysRevB.89.205414} and multi-layer transition metal dichalcogenides \cite{PhysRevLett.108.196802, PhysRevLett.121.026402, PhysRevLett.122.086402, tang_simulation_2020, regan_mott_2020, zhang_flat_2020, wang_correlated_2020, xu_correlated_2020}, where the moir\'e pattern leads to an enlarged unit cell, 
as well as in recent work examining moir\'e potential modulations on TI surface states \cite{wang2021moire,cano2020moire}. 

In this paper, inspired by these previous developments, we study magnetic skyrmion lattices on a TI surface and explore the 
resulting electronic states. Our work is also motivated by the desire to understand the interplay of the momentum space topology of TIs, as reflected in their helical Dirac surface states, with the topological real space texture of magnetic skyrmions. For instance, materials such as topological Kondo insulators (TKIs) can have Dirac surface states together with soft magnetic modes in the bulk due to strong correlation effects \cite{TKIreview}. Such materials might thus be prone to spontaneous magnetic ordering 
and time-reversal breaking at the surface \cite{Efimkin2014,Wolgast_Curbino_SMB6_2015,nakajima_SmB6_2016,2017_Tiwari_DWBS}.
The inversion breaking at the TKI surface can also enhance the role of chiral Dzyaloshinskii-Moriya magnetic exchange interactions \cite{IDM1,IDM2}, which 
could favor the formation of skyrmions at the surface. Magnetic topological materials such as MnBi$_2$Te$_4$ \cite{ChangLiu2019}
are another proposed candidate for realizing skyrmions \cite{McQueeney2020}.
Further possibilities of realizing magnetic skyrmions at TI surfaces include the ordering of impurity magnetic moments of dopants induced by RKKY interactions \cite{2009_Liu_RKKY}, proximity coupling to a magnetic substrate hosting these textures \cite{yasuda_geometric_2016,jiang_concurrence_2020,Li_TI_MI_2020,tokura2010,PhysRevB.98.060401}, or spontaneous magnetic ordering due to hexagonal warping of the surface Dirac cone \cite{Panagiotis}.

Previous theoretical work demonstrated the electrical charging of nonuniform magnetic textures, such as vortices and domain walls of N\'eel and Bloch type, by coupling to Dirac TI surface states \cite{Nagaosa2010}. Focusing on an isolated Bloch skyrmion texture, it was subsequently shown that chiral bound states, confined to the skyrmion perimeter and analogous to the Jackiw-Rebbi zero mode, offer a complementary mechanism for inducing electric charge \cite{2015_Hurst_Skyrmion_BS, JackiwRebbi}. Later studies investigated the scattering of Dirac electrons off of single skyrmions \cite{Araki, Wang_PRR}, and showed that the in-gap bound states modify the skyrmion-skyrmion interaction potential \cite{KunalSkyrmion}.

In ordinary magnetic metals, the presence of skyrmions is often inferred from an additional Hall contribution generated by the real-space Berry curvature induced by skyrmions. Termed the topological Hall effect (THE) \cite{THE1, THE2, PhysRevB.92.115417, PhysRevB.95.094413, PhysRevB.100.174411, QAHE_graphene_skyrmions}, this transport phenomenon is distinct from the anomalous Hall effect that derives from momentum-space Berry curvature \cite{Li_TI_MI_2020}. 
Recent experiments on magnetically doped TI superlattices 
have interpreted
anomalies in the Kerr effect \cite{Liu_2020} as arising
from skyrmions.
However, we emphasize that the 
experimental identification and disentanglement of the skyrmion
contribution from magnetic inhomogeneities 
can be difficult in practice, as showcased by recent work on SrRuO$_3$ films 
\cite{Hallanomaly1,Hallanomaly2,Hallanomaly3}. More importantly, 
as we discuss below, massless
Dirac fermions moving in a skyrmion background do not 
sense the topological charge of the skyrmions as a magnetic
flux and should not exhibit the THE.


This paper is organized as follows. In Section \ref{sec:continuum}, we begin with a band theoretic analysis of Dirac fermions coupled to the periodic Zeeman texture of the skyrmion lattice. This `nearly-free Dirac electron' approach allows us to determine the energy bands and their topological invariants, as well as symmetries. In Section \ref{sec:tight_binding}, we pass to a tight-binding description of the low-energy Bloch skyrmion bands, drawing inspiration from the bound states of the single-skyrmion problem when constructing Wannier orbitals. We conclude with a summary of important results, possible limitations, and a discussion of promising future directions.

\section{Continuum band theory}\label{sec:continuum}

\subsection{Single skyrmion and skyrmion crystal ansatze}\label{section:isolated}
The unit vector magnetization of an isolated 2D skyrmion centered at the origin 
may be written in the following form:
\begin{align}\label{eq:single_sk}
\bm{n}(r,\phi)=\begin{pmatrix}\sqrt{1-n_z(r)^{2}}\cos(\phi + \phi_{0})\\ \sqrt{1-n_z(r)^{2}}\sin(\phi + \phi_{0})\\ n_z(r) \end{pmatrix}.
\end{align}
We assume that $n_z(r)$ is a function which increases monotonically from the value $n_z(0)=-1$ at the skyrmion center to $n_z(r\geq R_0)=+1$ beyond a cutoff radius $R_0$. The fixed angle $\phi_0$ determines 
the skyrmion handedness. We highlight two special cases: 
``hedgehog''-type N\'eel skyrmions characterized by $\phi_0=0$, and ``vortex''-type Bloch skyrmions which have $\phi_0=\pm \pi/2$. The skyrmion topological charge 
\begin{align}\label{eq:Q}
Q_{\rm top}=-\frac{1}{4\pi} \int d^2{\bm r}~\bm{n}\cdot\partial_x \bm{n}\times\partial_y \bm{n}=1
\end{align}
is independent of $\phi_0$ and is invariant under local 
deformations of the texture.
We do not discuss the energetic stability of the various skyrmion types, but instead
present results for both Bloch and N\'eel skyrmions.

The radial function $n_z(r)$ may in general have a sharp, i.e. domain wall-like, or more gradual transition as a function of $r$. We parametrize this freedom by
\begin{align}\label{eq:transition_sin2}
n_{z}(r)=\begin{cases}
	-1 & r \in [0,\alpha R_0]\\
	2\sin^2\left(\frac{\pi(r-\alpha R_0)}{2(R_0-\alpha R_0)}\right)-1 & r\in(\alpha R_0, R_0)\\
	+1 & r \geq R_0 \end{cases}
\end{align}
In the limit $\alpha \!\to\! 1$, this ansatz leads to a sharp transition, with $n_z(r < R_0)=-1$ and $n_z(r > R_0)=+1$; in this case, the skyrmion approaches the form of a minority domain droplet with no in-plane magnetization component, and the distinction between N\'eel versus Bloch skyrmion loses its significance. On the other hand, the transition is smooth for all $\alpha<1,$ with the most gradual transition $n_z(r) = 2\sin^2(\pi r/2R_0)-1$ obtained when $\alpha=0.$ Tuning $\alpha\in[0,1)$ allows us to interpolate between these two limits, and we find that many of our results concerning the Chern numbers of the skyrmion bands depend crucially on this parameter. For later use, we also define the skyrmion core size by
\begin{equation}\label{eq:R_and_R0}
R = R_0 (1+\alpha)/2,
\end{equation}
which is where $n_z({r})$ undergoes a change in sign. This radius $R$ will later be found to determine the radius of ring-like Wannier functions obtained from the skyrmion bands and, more broadly, is a convenient tuning parameter for studying the skyrmion bands and their wavefunction topology. We remark that a variety of alternative forms for $n_{z}(r)$ in the interval $(\alpha R_0,R_0)$ have been investigated, but these lead to only minor quantitative differences as compared to Eq.~(\ref{eq:transition_sin2}). To simplify the discussion, we therefore focus on this particular form.

We construct the skyrmion crystal ansatz as a triangular lattice of skyrmions centered at Bravais vectors
\begin{equation}\label{eq:bravais_vectors}
    \bm{R} = m_1\bm{a}_1 + m_2\bm{a}_2,\quad m_{1,2}\in\mathbb{Z},
\end{equation}
where $\bm{a}_1=a(1,0)$ and $\bm{a}_2 = a\left(\frac{1}{2},\frac{\sqrt{3}}{2}\right)$. We assume for simplicity that $a > 2 R_0$, so that individual skyrmions in the crystal do not directly overlap, therefore allowing the magnetization $n_z(\bm{r})=+1$ of adjacent skyrmions to join smoothly at their Wigner-Seitz cell boundaries. Note that the spacing between skyrmions is controlled solely by the parameter $a,$ not by the quantities $R,R_0$ or $\alpha.$ We will denote the skyrmion reciprocal lattice by $\mathfrak{G}$ and its primitive vectors by
\begin{align}\label{eq:reciprocal_vectors}
    \bm Q_1 = Q(\sqrt{3}/2, -1/2),\quad \bm Q_2=(0,Q)
\end{align}
with $Q = 4\pi/\sqrt{3}a.$ The skyrmion lattice texture can be written as a Fourier series in these reciprocal lattice vectors, $\bm{n}(\bm{r}) = \sum_{\bm G \in\mathfrak{G}}\bm n_{\bm{G}} e^{i\bm r\cdot\bm G}.$

\begin{figure}[t]
    \subfloat{\includegraphics[width=0.49\textwidth]{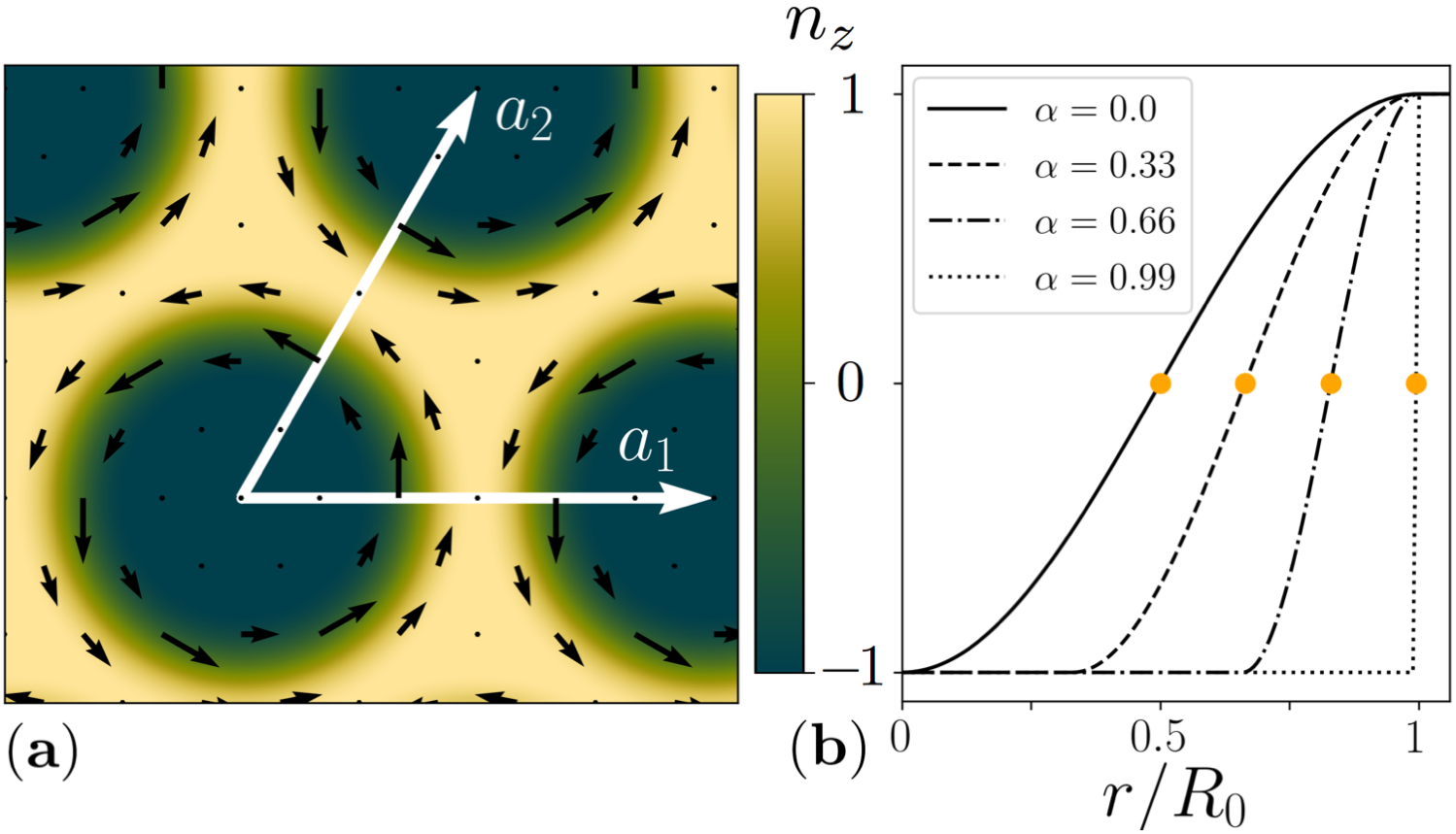}}
    \caption{\textbf{(a)} Bravais vectors $\bm{a}_1,\bm{a}_2$ and overheard view of the triangular lattice of Bloch skyrmions at $\alpha=0.5$ and cutoff radius $R_0/a=0.49$. The color corresponds to the out-of-plane magnetization $n_z$ whereas the vector field indicates the in-plane magnetization. \quad \textbf{(b)} The single-skyrmion magnetization profile $n_z$ plotted for various values of $\alpha$ which interpolates smoothly between the sinusoidal and domain wall limits. For each $\alpha$ we mark in orange the corresponding radius $R$ at which the out-of-plane magnetization changes sign.}
    \label{fig:profile_and_texture}
\end{figure}

\subsection{Coupling helical Dirac fermions to the magnetic
skyrmion crystal}

The continuum Hamiltonian for the TI Dirac surface states 
coupled to the skyrmion spin texture via a Hund's term is given by $H=H_0+H_1$, where
\begin{equation}\begin{aligned}
\label{eq:continuum_Ham}
    H_0 &= v_F \int_{\mathbb{R}^2} d^{2}\bm{r}\ c^{\dagger}(\bm{r})
    \left[\left(-i\hbar \frac{\partial}{\partial\bm{r}}\times\bm{\sigma}\right)\cdot \hat{z}\right] c(\bm{r}) \\
    H_1 &= J_{\rm eff} \int_{\mathbb{R}^2} d^{2}\bm{r}\ c^{\dagger}(\bm{r}) \bm{n}(\bm{r})\cdot\bm{\sigma} c(\bm{r}).
\end{aligned}\end{equation}
Here, the fermion operators are implicitly spinors. The quantities $v_F$ and $J_{\rm eff}$ denote the Dirac velocity and spin-fermion coupling strength, respectively. 

Here we argue that contributions from external magnetic fields may be neglected in various cases. First, we note that a non-vanishing Zeeman shift may be absorbed into the 
skyrmion texture without modifying any Hamiltonian symmetries. Thus, we only need to discuss the impact of an orbital magnetic field. In case where the Zeeman field is necessary to stabilize 
skyrmions, the corresponding orbital magnetic flux per skyrmion 
lattice unit cell is expected to be small 
compared to the flux quantum in the case of small skyrmions. 
For example, in $\text{MnBi}_2\text{Te}_4,$ the required magnetic field is on the order of $B\sim 0.02\text{ meV}/g\mu_B \sim 0.2\text{ T}$. This gives a magnetic length of $\ell_B\sim 60\text{ nm},$ so that the magnetic flux is small, $Ba^2 \sim \Phi_0 a^2 / 2\pi\ell_B^2 \ll \Phi_0$ with $a\sim 4\text{ nm}$ \cite{McQueeney2020}. Finally, we note that 
it may be possible to interface TI surface states with materials hosting skyrmions at zero-field, such as those recently reported to be stabilized by frustration \cite{Meyer2019} or soft X-ray illumination \cite{Schutz2020}.


Henceforth, we will measure energies in units of $\hbar v_F/a$. Let us denote the dimensionless spin-fermion exchange coupling by 
$J \!=\! (a/\hbar v_F) J_{\rm eff}$.
To estimate this in the
magnetic topological insulator MnBi$_2$Te$_4$,
we set
$\hbar v_F \!\sim\! 1\si{eV}\si{\angstrom}$ as
measured from angle resolved photoemission spectroscopy (ARPES) \cite{ChangLiu2019}. The spin-fermion coupling $J_{\rm eff}$
may be crudely estimated via $T_N \! \sim \! 
J^2_{\rm eff}/W$ within
an RKKY picture, where the bandwidth is $W \! \sim \! 1$\,eV
based on ARPES and the experimental 
N\'eel temperature is $T_N \! \sim\! 25$K \cite{ChangLiu2019}. 
This leads to 
$J_{\rm eff}\!\sim\! \sqrt{T_N W} \! \sim \! 50$\,meV. 
To estimate the skyrmion
lattice constant, we note that skyrmions in MnSi 
have $a \sim \! 20$\si{nm}, but MnBi$_2$Te$_4$ has been argued to
support skyrmions with a smaller lattice constant 
$\sim \! 4$\si{nm} \cite{McQueeney2020}. 
Taking
$a \sim \! 4$-$20$\si{nm} translates to a range of
a dimensionless spin-fermion couplings $J \! \sim \!1$-$10$.

Moving to momentum space and folding into the first Brillouin zone (BZ) of the skyrmion lattice,
\begin{equation}
    c_{s\bm G}(\bm{k})\equiv c_{s}(\bm{k}+\bm G) = \int_{\mathbb{R}^2} d^2\bm{r} e^{-i(\bm{k}+\bm{G})\cdot\bm{r}}c_s(\bm{r})
\end{equation}
the Hamiltonian is block-diagonal in the crystal momentum $\bm{k}$ due to discrete translational symmetry:
\begin{equation}
    H=\int_{\text{BZ}}\frac{d^{2}\bm{k}}{(2\pi)^{2}}\sum_{\bm{G},\bm{G}'\in\mathfrak{G}} c^{\dagger}_{\bm G}(\bm{k})\left(\mathscr{H}_{\bm{k}}\right)_{\bm{G},\bm{G}'}c_{\bm{G}'}(\bm{k}).
\end{equation}
In this basis, the Hamiltonian has matrix elements
\begin{align}\label{eq:H_qGG}
    \left(\mathscr{H}_{\bm{k}}\right)_{\bm{G},\bm{G}'}=
\delta_{\bm{G},\bm{G}'}\qty((\bm{k}+\bm{G})\times\bm{\sigma})_{z}-J\bm{n}_{\bm{G}-\bm{G}'}\cdot \bm{\sigma}
\end{align}
where the Pauli matrices correspond to spin. This defines for us the electronic band structure problem for the skyrmion crystal. Since the set of skyrmion reciprocal lattice vectors $\mathfrak{G}$ is infinite, each matrix $\mathscr{H}_{\bm{k}}$ possesses infinitely many components. Computation of the band structure therefore requires truncating $\mathfrak{G}$ to some finite number of reciprocal vectors nearest zero. This truncation is justified by the absence of any singularity in the skyrmion spin texture, corresponding to rapid decay of its Fourier components. For the range of parameters $0\leq J\lesssim 10$ we find that truncation to 300 momenta, and therefore 600 bands due to spin, is sufficient to attain convergence in the energies, Berry curvature, and tight-binding parameters considered later.

The Bloch vectors $u_{s\bm G,n}(\bm{k})$ are defined as the eigenvectors of $\mathscr{H}_{\bm{k}}$ and allow us to define the Bloch operators
\begin{equation}\label{eq:Bloch_expansion}
    \psi_{\bm{k} n}^{\dagger}=\sum_{s\bm G}c_{s\bm G}^{\dagger}(\bm{k})u_{s\bm G,n}(\bm k).
\end{equation}
By construction, the second-quantized Hamiltonian is diagonal in the band basis,
\begin{equation}\label{eq:continuum_Ham_diagonal}
    H = \int_{\text{BZ}}\frac{d^2\bm{k}}{(2\pi)^2} \sum_{n} \epsilon_{n}(\bm k)\psi_{\bm{k}n}^{\dagger}\psi_{\bm{k}n}.
\end{equation}

Before we separately present results for the band structure for Bloch and N\'eel skyrmions, we note that the skyrmion crystal reduces the continuous rotational symmetry of the isolated skyrmion problem to a six-fold symmetry. We fix the rotation axis to be parallel to $\hat{z},$ passing through a skyrmion center. We represent the six-fold operator on the continuum states by
\begin{equation}\label{eq:C6}
    C_6 c(\bm r) C_6^\dagger = e^{i \frac{\pi}{3} \frac{\sigma_z}{2}} c(C_6 \bm r),
\end{equation}
where $C_6$ acts as a six-fold counterclockwise rotation on vectors. By appeal to the identity $C_6\bm{n}(C_6^{-1}\bm{r})=\bm{n}(\bm{r}),$
one finds that $C_6$ is a symmetry (see Supplemental Material \cite{SuppMat}), independent of the value of $\phi_0$ which sets the skyrmion type.

\section{Bloch skyrmion lattice}

\subsection{Continuum model}

When the magnetization texture $\bm{n}(\bm{r})$ describes a lattice of Bloch skyrmions, given by setting $\phi_0=\pi/2$ or $-\pi/2,$ we are granted several additional symmetries which constrain the energy bands and Berry curvature. Setting $\phi_0$ accordingly in Eq.~(\ref{eq:single_sk}), one observes that the magnetization of an isolated Bloch skyrmion satisfies the property $\bm r\cdot \bm n(\bm r)=0.$ Consequently, the in-plane divergence must vanish identically, $\nabla^\mathrm{2D}\cdot\bm n(\bm r)=0$ due to the divergence theorem \cite{2015_Hurst_Skyrmion_BS}. In this case, it has been shown \cite{Araki} that the in-plane component of $\bm{n}(\bm{r})$ may be entirely removed from the Hamiltonian by the gauge transformation $\mcU(\bm{r})=\exp\left(iJ \int_0^{\rho}d\rho'\sqrt{1-n_{z}^{2}(\rho')}\right).$ All of these statements remain true when we promote the texture to a triangular lattice of skyrmions rather than an isolated skyrmion. One must simply interpret $\bm{r}=(\rho,\phi)$ in Eq.~(\ref{eq:single_sk}) as a quantity measured relative to the nearest skyrmion center.

The fact that the symmetry $\mcU(\bm{r})$ is periodic in the skyrmion lattice ensures that the band structure is invariant under its action. This is because it performs a unitary convolution on each subspace of definite crystal momentum. Therefore, we may drastically simplify the arguments that follow by setting the in-plane magnetization to zero outright. We denote the resulting texture by a separate symbol to distinguish it from the original periodic texture:
\begin{align}
    \mathbbm{n}(\bm r) \equiv (0,0,n_z(\bm r)).
\end{align}
Note that $\mathbbm{n}(\bm{r})$ is neither normalized nor does it possess a winding number. Moreover, the texture crucially possesses the symmetries
\begin{align}\label{eq:twofold_PH}
    \mathbbm{n}(\bm r) = \mathbbm{n}(-\bm r) = \mathbbm{n}(M_y \bm r)
\end{align}
with $M_y$ defined as the matrix which flips the second component of a vector. We now proceed with a description the symmetries manifesting in the continuum description of the Bloch skyrmion lattice problem.

\subsubsection{Particle-hole symmetry}\label{sec:PH_symmetry}
Consider the unitary transformation which flips spins and exchanges electrons with holes at fixed momentum,
\begin{align}\label{eq:PH}
    U c(\bm{k}) U^\dagger = c^\dagger(\bm{k})\sigma_y,\quad UiU^\dagger = i.
\end{align}
Leveraging the identity $\mathbbm{n}(\bm r)=\mathbbm{n}(-\bm r),$ we demonstrate (see Supplemental Material \cite{SuppMat}) that $U$ is a symmetry of the Hamiltonian, satisfying $[U,H]=0.$ Such a symmetry is commonly referred to as a `particle-hole' symmetry \cite{tenfoldway}. Analyzing this symmetry in reciprocal space reveals that the band structure is symmetric about zero energy at each value of the crystal momentum. Indeed, in the notation of Eq.~(\ref{eq:H_qGG}) we find that the particle-hole symmetry is expressed by the matrix relation
\begin{align}\label{eq:matrix_relation_PH}
    \sigma_y \qty(\mathscr{H}_{\bm{k}})^* \sigma_y = -\mathscr{H}_{\bm{k}}.
\end{align}
Since $\mathscr{H}_{\bm{k}}$ is Hermitian then its spectrum is invariant under both complex conjugation and change of basis. The above equation thus implies that its spectrum is symmetric, i.e. that its eigenvalues come in pairs $\epsilon_{-n}(\bm{k})= - \epsilon_{n}(\bm{k}).$ We have adopted a convention where the valence (conduction) bands are labeled by negative (positive) integers in order of their energy.

\begin{figure}[t]
    \includegraphics[width=0.47\textwidth]{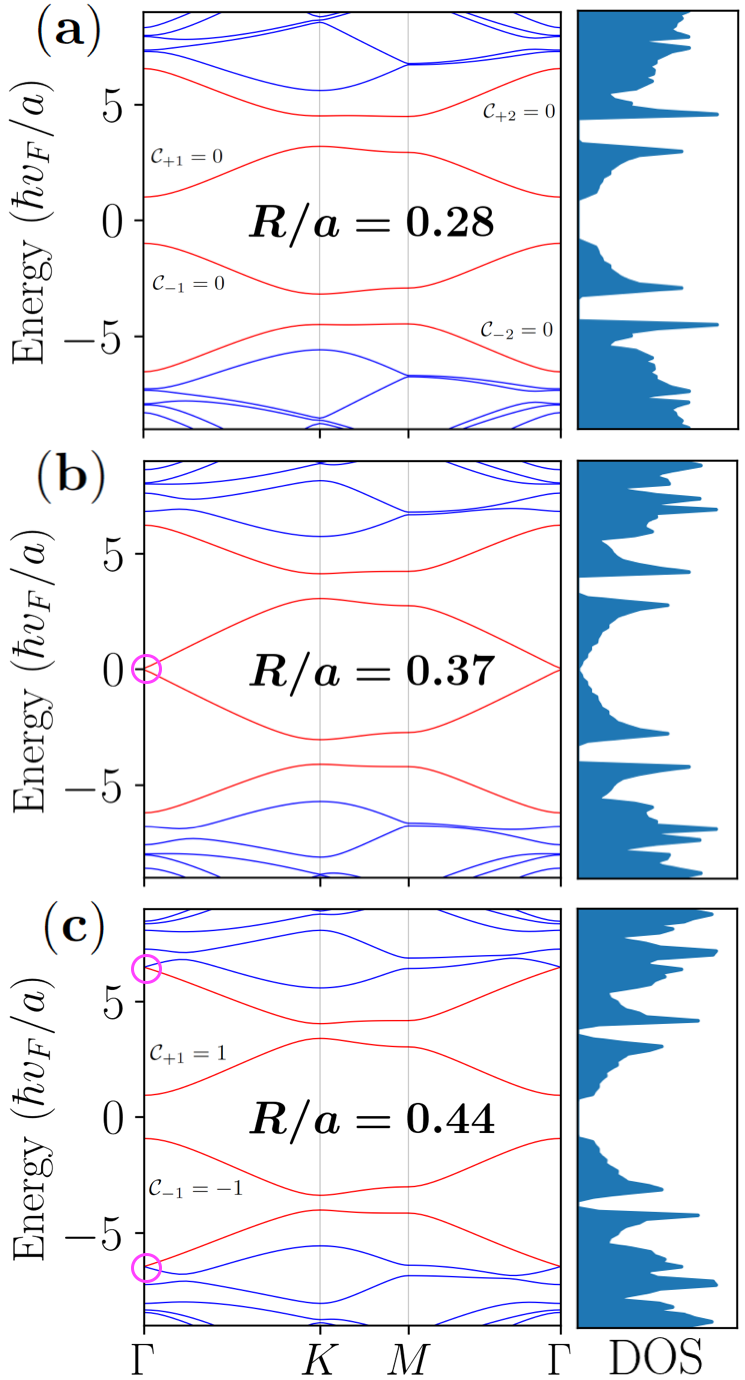}
    \caption{Bloch skyrmion bands and density of states (DOS) at $J=3$ and cutoff radius $R_0/a=0.49.$ \textbf{(a)} At small $R/R_0,$ the four lowest-energy bands carry zero Chern number and we observe windows of vanishing DOS. \textbf{(b)} As the core size $R$ is increased there is a gap closure (circled) between the $n=\pm1$ bands at the $\Gamma$ point, resulting in Chern numbers $\mathcal{C}=(0,-1\!:\!1,0)$. The dispersion and DOS at this transition are linear. \textbf{(c)} As $R$ is increased towards its maximum value $R_0,$ the bands undergo a final transition into the $\mathcal{C}=(1,-1\!:\!1,-1)$ sector upon direct gap closure between the $|n|=2,3$ bands at the $\Gamma$ point. Increasing $J$ leads to narrower Chern bands and more pronounced peaks in the DOS.}
    \label{fig:bloch_J=3_bands}
\end{figure}

\subsubsection{Chiral symmetry}\label{sec:chiral_symmetry}

Continuing with our analysis of the Bloch skyrmion lattice, we now show that the Hamiltonian Eq.~(\ref{eq:continuum_Ham}) satisfies a `chiral' symmetry \cite{tenfoldway} which constrains the band structure and Berry curvature. Consider the anti-unitary operator which flips spin and exchanges electrons with holes at mirror-related momenta,
\begin{align}\label{eq:chiral}
    A c_{\bm{G}}(\bm{k})A^\dagger = c^\dagger_{M_y \bm{G}}(M_y \bm{k})\sigma_x,\qquad AiA^\dagger = -i.
\end{align}
At the level of the Hamiltonian kernel, the symmetry $[A,H]=0$ derives from the identity
\begin{align}\label{eq:My_matrix_conjugation}
    \qty(\mathscr{H}_{\bm{k}})_{\bm{G},\bm{G}'} = - \qty(\sigma_x\,\mathscr{H}_{M_y\bm{k}}\,\sigma_x)_{M_y\bm{G},M_y\bm{G}'}.
\end{align}
For each eigenvector $u_n(\bm{k})$ of $\mathscr{H}_{\bm{k}},$ we therefore have a related eigenstate $u_{-n}(M_y\bm{k})$ of $\mathscr{H}_{M_y\bm{k}}$ with opposite energy. As shown explicitly in the Supplemental Material \cite{SuppMat}, this provides a relation between the Berry curvature of the opposing $\pm n$ bands at these mirror-related momenta,
\begin{align}\label{eq:Berry_constraint}
    F^{(n)}(\bm{k}) = -F^{(-n)}(M_y\bm{k})
\end{align}
Upon integrating over the Brillouin zone, we immediately see that these bands must carry opposite Chern number, $\mathcal{C}_{-n} = -\mathcal{C}_n.$ As we later discuss, this property is absent from the N\'eel band structure.

\subsubsection{Topological bands and DOS}\label{sec:Bloch_bands}

In Fig.~\ref{fig:bloch_J=3_bands} we illustrate the band structure for the continuum Hamiltonian Eq.~(\ref{eq:continuum_Ham}) in the Bloch skyrmion case. In accordance with the particle-hole symmetry presented in Eq.~(\ref{eq:matrix_relation_PH}), the energy spectrum is symmetric everywhere in the skyrmion Brillouin zone. We exhibit the bands for fixed parameter values $J=3$ and $R_0/a=0.49,$ with the latter chosen so as to encourage hybridization between the single-skyrmion bound states. We recall that $R_0$ sets the cutoff radius at which the skyrmion magnetization is purely polarized in the $\hat{z}$ direction whereas the core size $R,$ as illustrated in Fig.~\ref{fig:profile_and_texture}, sets the radius at which the Dirac mass $n_z$ changes sign.

At $R/a=0.28$ the four bands nearest half-filling each carry zero Chern number and are continuously connected to an atomic insulator phase in the limit of large $J.$ When the skyrmion core size $R/a$ is increased up to $0.37$ we observe a gap closure between the $n=\pm1$ bands at the $\Gamma$ point upon which the system enters a new topological sector with Chern numbers $\mathcal{C}=(0,-1\!:\!1,0)$ for the four bands nearest-half-filling. As $R$ is increased toward its maximal value $R_0/a=0.49,$ we reach the value $R/a=0.44$ at which a final topological transition takes us into the $\mathcal{C}=(1,-1\!:\!1,-1)$ sector, this time due to a gap closure at the $\Gamma$ point between the $|n|=2,3$ bands. Unlike in a related moir\'e study \cite{cano2020moire}, these remote Dirac cones are not protected by time-reversal, nor can their velocity be easily tuned to zero within the present model.

It is important to distinguish this ``intrinsic" anomalous Hall effect, permitted by broken time-reversal symmetry, from the well-studied THE which arises when nonrelativistic electrons are coupled to a skyrmion lattice \cite{THE1,THE2}. 
The THE is absent for Dirac electrons
since the effective magnetic flux density $\bm{B}_\text{eff} \sim \nabla^\mathrm{2D}\cdot\bm{n}$
seen by the Dirac electron in a skyrmion texture
does not depend on the topological charge Eq.~(\ref{eq:Q}) 
of the magnetic texture \cite{2017_Han}. Indeed, this 
effective flux density, when
integrated over the skyrmion unit cell, vanishes for any generic skyrmion texture due to the divergence theorem (for any type of skyrmion). For Bloch skyrmions, an even stronger condition holds, that $\bm{B}_\text{eff}$ itself vanishes everywhere 
in space.

\begin{figure}[t]
    \subfloat{\includegraphics[width=0.42\textwidth]{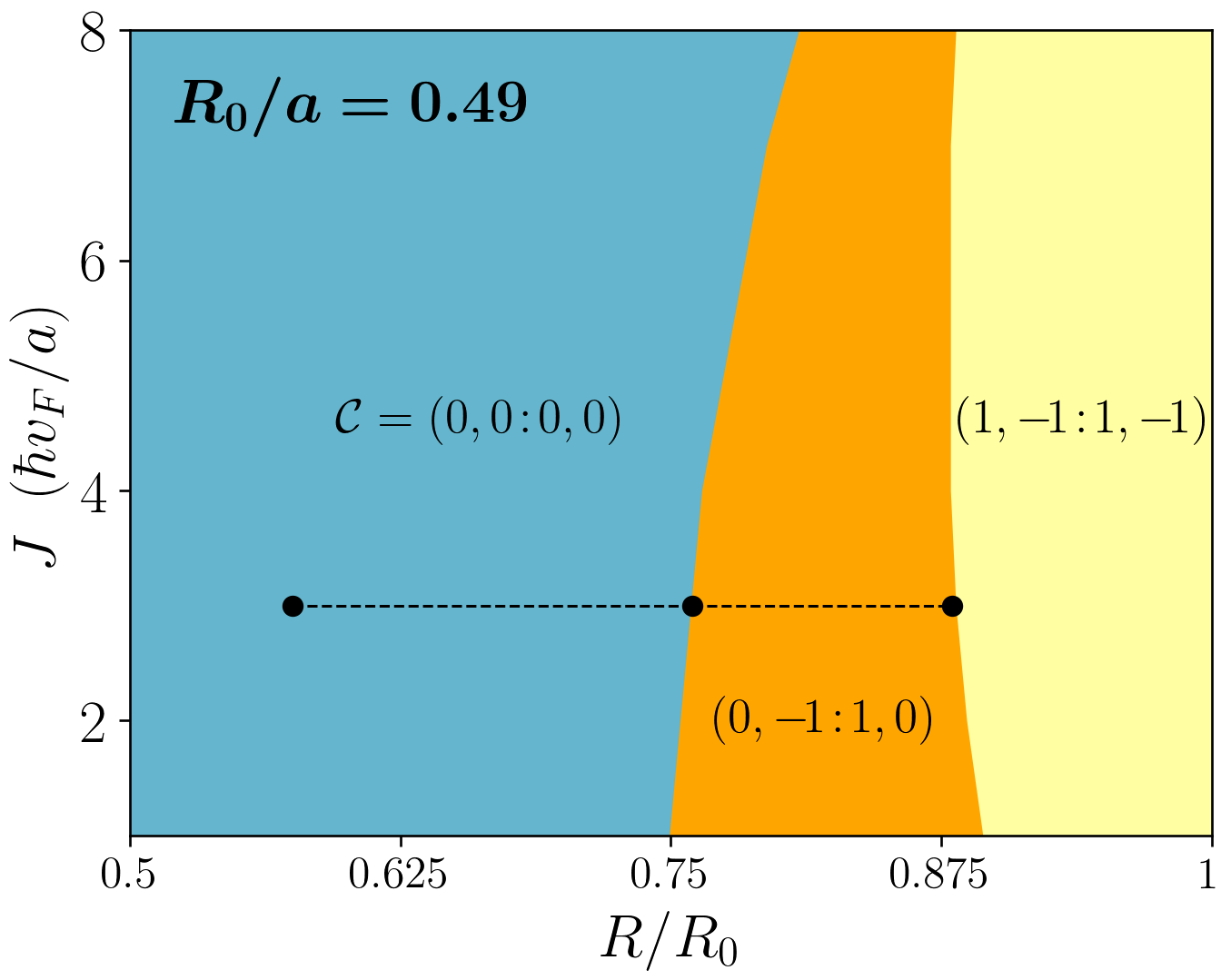}}
    \caption{Chern number phase diagram for the four lowest-energy bands in the Bloch skyrmion case. The chosen value $R_0/a=0.49$ corresponds to a near-maximally dense packing of skyrmions. The Chern numbers in each tuplet $\mathcal{C}$ are listed in order of increasing energy with a colon marking half-filling. In agreement with the Berry curvature constraint Eq.~(\ref{eq:Berry_constraint}), in each sector we observe that the Chern numbers of bands of opposite energy are opposite, $\mathcal{C}_{-n} = -\mathcal{C}_n.$ For ease of comparison, we mark points corresponding to the band structures presented in Fig.~\ref{fig:bloch_J=3_bands}.}
    \label{fig:Bloch_chern_phase_diagram}
\end{figure}

The symmetry $\mathcal{C}_{-n}=-\mathcal{C}_n$ of these Chern numbers is consistent with the chiral symmetry constraint, Eq.~(\ref{eq:Berry_constraint}). In each case, the Chern numbers were computed from the band eigenstates following the methodology and sign conventions of \cite{Fukui_Chern}. In Fig.~\ref{fig:Bloch_chern_phase_diagram} we display a phase diagram indicating the Chern numbers carried by the four lowest-energy bands over the range of parameters $J\in[1,8]$ and $R\in[R_0/2,R_0]$ at fixed $R_0/a=0.49.$ We omit the range $J\in[0,1]$ where the bands too closely resemble those of the free Dirac Hamiltonian. We observe that all horizontal cuts along the phase diagram realize the same topological phases. To aid in visualizing the appearance of the corresponding bands, we mark in this figure those values of $R$ whose bands are displayed in Fig.~\ref{fig:bloch_J=3_bands}.

As the cutoff radius $R_0$ is decreased, the $\mathcal{C}=(1,-1\!:\!1,-1)$ region recedes completely, followed by the $\mathcal{C}=(0,-1\!:\!1,0)$ phase. Once the cutoff radius has reached $R_0=0.35a,$ all four bands carry zero Chern number over the entire range of parameters $(R,J)$. This phase can be understood as being continuously connected to the limit of well-separated Wannier orbitals which resemble the electronic bound states of isolated skyrmions. On the other hand, fixing $R_0/a\sim0.49$ as in Fig.~\ref{fig:Bloch_chern_phase_diagram} and increasing the skyrmion core size $R$ toward $R_0,$ the domain wall limit, we find that increasing $J$ leads to the $|n|=1,2$ bands bunching together and flattening, reminiscent of the single-particle physics of magic angle bilayer graphene \cite{Bistritzer_TBG}. In the trivial Chern sector, increasing $J$ more simply increases the number of in-gap bands, which flatten and are completely isolated.

\subsection{Tight-binding model for the Bloch skyrmion lattice problem}\label{sec:tight_binding}
We can approach the problem of Dirac electrons coupled to a Bloch skyrmion texture from a complementary angle, namely a tight-binding approach. In the previous sections we argued that the continuum problem Eq.~(\ref{eq:continuum_Ham}) features both a particle-hole symmetry, responsible for a symmetric energy spectrum, and a chiral symmetry, which dictates that the particle-hole-related bands must carry opposite Chern number. Well-known theoretical results about the interplay between Wannier functions and band topology guarantee the existence \cite{Wannier_existence} of a Wannier representation for any even number of non-degenerate bands nearest half-filling. In the limit of large interskyrmion distance $a \gg R_0$, we expect the Wannier orbitals to approach single-skyrmion electronic bound states, whose features we review in the following section.
Application of the Wannier projection method \cite{MLWF_1997,MLWF_RMP} for the two gapped bands nearest half-filling, which we discuss in Sec.~\ref{sec:Wannier}, reveals well-localized Wannier states whose qualitative features match those of the single-skyrmion bound states.

\subsubsection{Single-skyrmion bound states}\label{subsc:single_skyrmion}
The problem of Dirac fermions coupled to a single skyrmion was addressed by Hurst {\it et al.} in Ref.~[\onlinecite{2015_Hurst_Skyrmion_BS}]. In that work, the authors considered a two dimensional Dirac model with a position-dependent Dirac mass representing the skyrmion. The corresponding wavefunctions were found to be strongly peaked at the skyrmion perimeter where the Dirac mass changes sign. Further studies \cite{KunalSkyrmion} found similar states for a more realistic description of the skyrmions. Most importantly, the skyrmion bound states are a discrete set of states with energies inside the bulk electronic gap and a well-defined out of plane angular momentum $j=\pm 1/2,\pm 3/2,\ldots$ arising from the rotational symmetry of the single skyrmion texture. The bound state wavefunctions take the form
\begin{eqnarray}\label{eq_totalwavefn}
\Psi_j(\bm{r}) = \begin{pmatrix}
        e^{i(j-\frac{1}{2})\phi}\chi_{j,\uparrow}(r)\\e^{i(j+\frac{1}{2})\phi}\chi_{j,\downarrow}(r)
    \end{pmatrix}.
\end{eqnarray}
The asymptotic behavior of the radial wavefunctions $\chi_{j,s}(r)$ has been extensively studied \cite{2015_Hurst_Skyrmion_BS, KunalSkyrmion, Araki}. While their exact form is not important for our purposes, they are known to decay exponentially at long distances.

\subsubsection{From Bloch states to Wannier states}
Combining Bloch skyrmions together in a hexagonal lattice, we expect the skyrmion bound states to hybridize to form orthogonal Wannier states. Supposing we've identified some set of isolated bands
\begin{equation}\label{eq:active_bands}
    n\in\{\pm 1,\pm2,\dots, \pm n_0\} \equiv \mathcal{B}    
\end{equation}
nearest half-filling, the Wannier states are merely the Fourier transform of a smoothening of the Bloch states
\begin{equation}\label{eq:wannier_fourier}
    d^\dagger_{\bm{R}j} = \frac{1}{\sqrt N}\sum_{\bm{k}} d^\dagger_{\bm{k}j} e^{-i\bm{k}\cdot\bm{R}}
\end{equation}
for a choice of unitaries $\mathscr{U}_{nj}(\bm{k})$ such that the rotated states
\begin{equation}\label{eq:wannier_to_bloch_unitary}
    d^\dagger_{\bm{k}j} = \sum_{n\in\mathcal{B}}\psi^\dagger_{\bm{k}n}\mathscr{U}_{nj}(\bm{k})
\end{equation}
are smooth in the variable $\bm{k}.$ Less abstractly, these orbitals correspond to smooth functions $d_{s\bm{k}j}(\bm{r})= \langle\Omega|c_s(\bm{r})d^\dagger_{\bm{k}j}|\Omega\rangle.$ In Sec. \ref{sec:Wannier} we implement a technique, known as the projection method \cite{MLWF_1997,MLWF_RMP}, for constructing the unitaries $\mathscr{U}_{nj}(\bm{k})$ directly from an initial guess for the Wannier orbitals. In particular, the constructed Wannier orbitals will carry an eigenvalue $j$ under $C_6,$ justifying the choice of $j$ as a label.

In a tight-binding representation of a manifold of bands $\mathcal{B}$, the Hamiltonian data is encoded in a set of amplitudes
\begin{equation}\label{eq:hopping}
    t_{\bm{\delta}jj'} = \langle\bm{\delta}\!+\!\bm{R}\,j|H|\bm{R}\,j'\rangle
\end{equation}
where $\bm{\delta},\bm{R}$ are skyrmion Bravais vectors. The independence of the hoppings on $\bm{R}$ is due to discrete translational symmetry. Since the Hamiltonian doesn't mix states with different angular momentum then the on-site overlap matrix $t_{\bm{0}jj'}\equiv \varepsilon_j \delta_{jj'}$ is diagonal. The orthogonal projection of the continuum Hamiltonian Eq.~(\ref{eq:continuum_Ham_diagonal}) into the bands $\mathcal{B}$ then reads
\begin{equation}
H_{\mathcal{B}} = \sum_{\bm{k}\in\mathrm{BZ}}\sum_{jj'} d_{\bm{k}j}^\dagger H_{\bm{k} j j'}\, d_{\bm{k}j'}
\end{equation}
with momentum space kernel given by
\begin{equation}\label{eq:Ham_kernel}
H_{\bm{k} j j'} \equiv \delta_{j j'} \varepsilon_j+\sum_{\bm{\delta}} e^{-i \bm{k} \cdot {\bm \delta}} t_{\bm{\delta} j j'}.
\end{equation}

In the following, we discuss how the rotational, particle-hole, and chiral symmetries act on the Wannier states as well as the constraints they impose on the hopping parameters. We focus here on nearest-neighbor hoppings, leaving a detailed analysis of next and next-next-nearest neighbor hoppings to the Supplemental Matterial \cite{SuppMat}. In the nearest-neighbor case, the displacement vector $\bm{\delta}$ runs over the six nearest neighbors
\begin{equation}
    \bm{\delta} = \pm \bm{a}_1, \pm \bm{a}_2, \pm \bm{a}_3
\end{equation}
where $\bm{a}_{1,2}$ are defined in Eq.~(\ref{eq:bravais_vectors}) and $\bm{a}_3=\bm{a}_2-\bm{a}_1.$ At this stage we do not limit our analysis to any particular number of bands $|\mathcal{B}|.$

\subsubsection{Representing the six-fold symmetry}\label{discrete-rotation-symmetry}

While the lattice of skyrmions breaks the continuous rotation symmetry of the single skyrmion texture, the $C_6$ of Eq.~(\ref{eq:C6}) still remains. We choose to implement it on the Wannier orbitals by
\begin{equation}\label{eq:C6_wannier_rep}
    C_6 d_{\bm{R}j} C_6^\dagger = e^{i\frac{\pi}{3}j}\,  d_{C_6\bm{R}\,j}
\end{equation}
which conveniently leads to Wannier functions which transform under $C_6$ like the single-skyrmion bound states, Eq.~(\ref{eq_totalwavefn}). Moreover, the hoppings are constrained (see Supplemental Material \cite{SuppMat}) to satisfy
\begin{equation}\label{rotation-transform}
t_{C_6\bm{\delta}\, j j'} = e^{-i\frac{\pi}{3}(j\! -\! j')} t_{\bm{\delta} j j'}.
\end{equation}
This property allows us to express all the nearest neighbor hopping parameters in terms of $t_{jj'}\equiv t_{\bm{a}_1 j j'}$ alone. Correspondingly, Eq.~(\ref{eq:Ham_kernel}) takes the compact form
\begin{equation}\label{eq:wannier_k_hamiltonian}
H_{\bm{k} j j'} = \delta_{j j'} \varepsilon_j + s_{\bm{k} j j'} t_{j j'}
\end{equation}
where $s_{\bm{k} j j'},$ a function of the nearest-neighbor lattice geometry whose precise form is inessential to the present discussion, may be found in the Supplemental Material along with the generalization to further-neighbors \cite{SuppMat}.

For later reference, we remark that Hermiticity and $C_2$, whose action is obtained by three applications of the $C_6$ constraint Eq.~(\ref{rotation-transform}), together guarantee
\begin{equation}\label{rotation}
t_{j'j} = e^{-i\pi(j\!-\!j')}\left( t_{j j'} \right)^*.
\end{equation}
Finally, while our tight-binding construction is nominally performed for the Bloch case for topological reasons, we remark that these $C_6$ results apply equally well to the N\'eel case provided that one is modeling a gapped subset of bands with net zero Chern number.

\subsubsection{Representing the particle-hole symmetry}\label{particle-hole-symmetry}
The particle-hole and chiral symmetries hold only in the Bloch case. We represent the former on the Wannier states by
\begin{equation}\label{eq:U_wannier_rep}
U d_{\bm{k}j} U^\dagger = \left(d_{\bm{k}\,-\!j} \right)^\dagger e^{i \pi j}.
\end{equation}
By equating the band-projected Hamiltonian $H_{\mathcal{B}}$ to the particle-hole conjugated expression
\begin{equation*}
U H_{\mathcal{B}} U^\dagger = -\sum_{\bm{k}\, j\, j'} d_{\bm{k}j}^\dagger\left[ H_{\bm{k} -\!j' -\!j} \, e^{i\pi(j'-j)}\right] d_{\bm{k}j'} + \sum_{\bm{k} j} H_{\bm{k} j j},
\end{equation*}
we immediately deduce that
\begin{equation}
H_{\bm{k} -\!j' -\!j} \, e^{i\pi(j'-j)} = - H_{\bm{k} j j'}.
\end{equation}
In the Supplemental Material \cite{SuppMat} we show that this condition is satisfied if and only if
\begin{equation}
    \begin{aligned}\label{eq:energy-and-particle-hole}
    \varepsilon_{-j} &= - \varepsilon_j \\
    t_{-\!j'\, -\!j} &= - e^{i\pi(j\!-\!j')} t_{j j'}
    \end{aligned}
\end{equation}

\subsubsection{Representing the chiral symmetry}\label{chiral-symmetry}
The final constraints on the hopping parameters derive from the chiral symmetry Eq.~(\ref{eq:chiral}), which we represent on the Wannier orbitals by
\begin{equation}\label{eq:A_wannier_rep}
A d_{\bm{k}j} A^\dagger = \left( d_{M_y\bm{k}\,-\!j} \right)^\dagger.
\end{equation}
As in our particle-hole analysis, comparison of $H_{\mathcal{B}}$ to
\begin{equation}
A H_{\mathcal{B}} A^\dagger = - \sum_{\bm{k}\, j\, j'} d_{\bm{k}j} ^\dagger\left[H_{M_y\bm{k}\, -\!j -\!j'}\right]d_{\bm{k}j'} + \sum_{\bm{k} j} H_{\bm{k} j j}
\end{equation}
demands immediately that
\begin{equation}
H_{M_y\bm{k}\, -\!j -\!j'} = - H_{\bm{k} j j'}.
\end{equation}
In the Supplemental Material \cite{SuppMat} we show that this condition is satisfied if and only if
\begin{equation}\begin{aligned}
\varepsilon_{-j} &=  - \varepsilon_j \\
t_{-\!j\, -\!j'} &= - t_{j j'} \label{chiral_hoppings}
\end{aligned}\end{equation}
Together with Eq.~(\ref{rotation}) and (\ref{eq:energy-and-particle-hole}), these expressions constitute all independent constraints on the parameters $\varepsilon_j$ and $t_{jj'}$ due to the symmetries of the Bloch skyrmion lattice system.

\subsection{Two-band Wannier construction}
We now perform an explicit construction of Wannier functions for the gapped low-energy bands of the Bloch skyrmion problem. We treat the two bands of lowest energy, $n\in \mathcal{B}=\{\pm 1\}.$ In our analysis of the continuum theory, these bands were found to maintain a finite energy gap to the remaining higher-energy states over the entire phase diagram Fig.~\ref{fig:Bloch_chern_phase_diagram}, with a gap closure occurring only mutually between them at zero energy.

The resulting tight-binding model features two Wannier states, labeled by their out-of-plane angular momentum $j=\pm1/2,$ and localized at each skyrmion site. Their wavefunctions are found to be tightly concentrated at the radius $R$ and exponentially decaying at long distances, which is reminiscent of the single-skyrmion bound states discussed in Sec. \ref{subsc:single_skyrmion}, even for a reasonably closely packed skyrmion lattice.

Finally, we numerically observe that the nearest-neighbor hopping amplitudes are generically dominant. This inspires the independent study of a nearest-neighbor two band toy model whose Chern sectors can be characterized analytically as a function of the hopping amplitudes $t_{{-\!{1/2}},{+\!{1/2}}}$ and $t_{{-\!{1/2}},{-\!{1/2}}}$ of Eq.~(\ref{eq:wannier_k_hamiltonian}). We find that the hopping amplitudes realized by the continuum model account for only half of the topological phases present in the two band toy model, thereby motivating further study into variants of Eq.~(\ref{eq:continuum_Ham}) as a means of realizing novel Chern insulator phases.

\subsubsection{Wannier orbitals of Bloch skyrmion bands}\label{sec:Wannier}

\begin{figure*}
    \centering
    \includegraphics[width=1.0\textwidth]{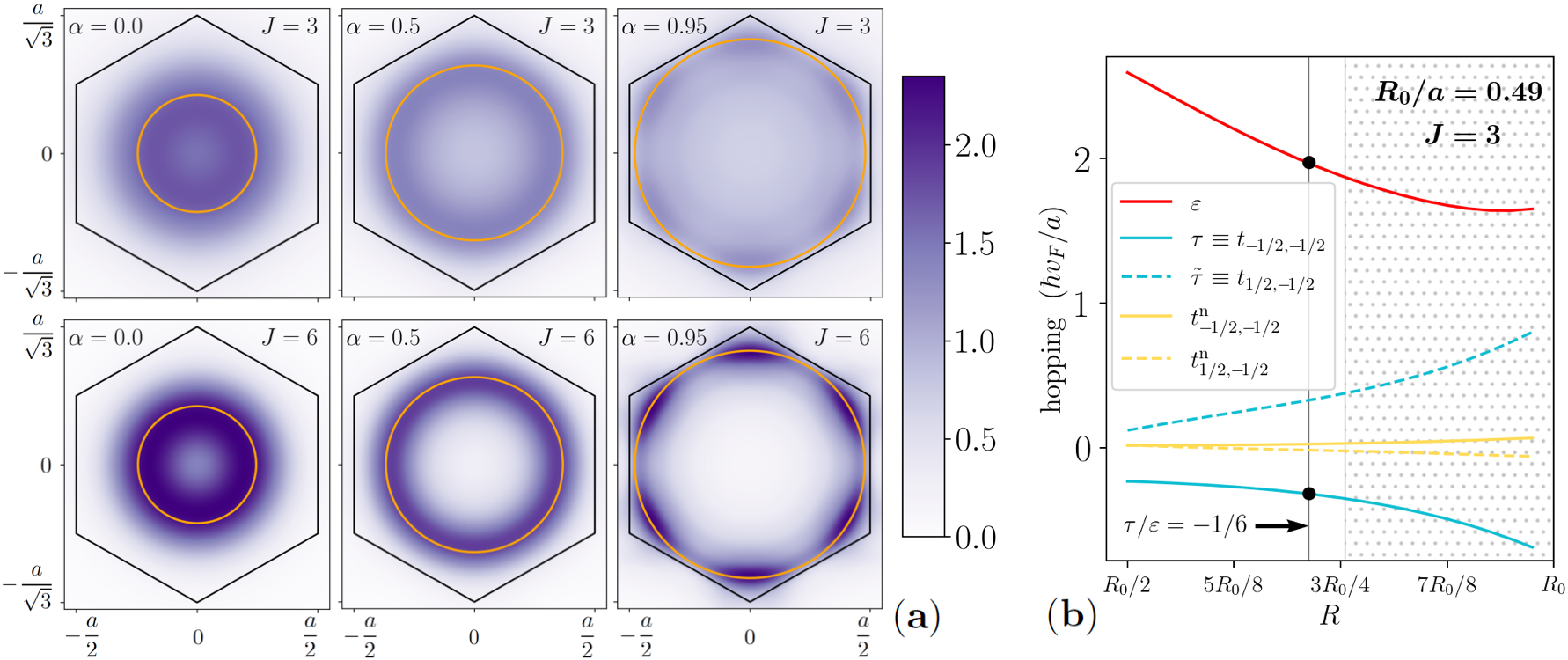}
    \caption{\textbf{(a)} Wannier function densities $\sum_s |d_{s\bm{0}j}(\bm{r})|^2$ derived from the lowest two Bloch skyrmion bands at $R_0/a=0.49.$ The density is equal for opposite $j.$ The black hexagon indicates the boundary of a Wigner-Seitz cell of the skyrmion lattice and the orange ring indicates the skyrmion core size, $R=R_0(1+\alpha)/2.$ The Wannier function density accumulates at $R$ and is more tightly localized for larger Hund's coupling $J.$ For small skyrmion size, the Wannier functions closely resemble the single skyrmion bound states. For larger skyrmion radius $R$ (i.e., larger $\alpha$), the lower six-fold symmetry of the Wannier function density is more apparent. \quad \textbf{(b)} Hopping parameters derived from the Wannier functions at $(J,R_0/a)=(3,0.49).$ The nearest-neighbor hoppings $\tau,\tilde\tau$ are most prominent. The shaded region marks the transition from the $\mathcal{C}=(0:0)$ sector to $\mathcal{C}=(-1:1),$ whereas the vertical line at $\tau/\varepsilon=-1/6$ marks where this boundary shifts upon truncating the hoppings at nearest-neighbor.}
    \label{fig:wannier_functions}
\end{figure*}

To construct the $j=\pm1/2$ Wannier states from the pair $n=\pm1$ of lowest-lying energy bands, we employ the projection method \cite{MLWF_1997,MLWF_RMP}. One begins with a set of trial orbitals $g_{sj}(\bm{r})$ which serve as a best guess for the Wannier orbitals centered in the home unit cell $\bm{R}=\bm{0}$. These trial states are then projected into the manifold of Bloch states at each wavevector $\bm{k},$
\begin{equation}
    |\phi_{\bm{k}j}\rangle = \sum_{n=\pm1}|\psi_{\bm{k}n}\rangle\langle \psi_{\bm{k}n}|g_j\rangle
\end{equation}
where the latter inner product is taken over all of space. To ensure that these states are orthonormal, one performs a Löwdin transformation
\begin{equation}
    |d_{\bm{k}j}\rangle = \sum_{j'} |\phi_{\bm{k}j'}\rangle (\mathscr{S}_{\bm{k}}^{-1/2})_{j'j}
\end{equation}
where the overlap matrix is given by
\begin{equation}
    \mathscr{S}_{\bm{k}} = \mathscr{A}^\dagger_{\bm{k}}\mathscr{A}_{\bm{k}},\qquad (\mathscr{A}_{\bm{k}})_{nj}\equiv\langle \psi_{\bm{k}n}|g_j\rangle.
\end{equation}
The Bloch-like states $|d_{\bm{k}j}\rangle$ will be smooth in $\bm{k},$ and the corresponding Wannier orbitals Eq.~(\ref{eq:wannier_fourier}) will be exponentially-localized, provided that the overlap matrix is finite everywhere in the BZ. As an immediate consequence of the above definitions, we obtain the unitary transformation taking us between the Bloch-like states $d_{\bm{k}j}$ and the original Bloch states. In the notation of Eq.~(\ref{eq:wannier_to_bloch_unitary}), the unitary is given by
\begin{equation}
    \mathscr{U}_{nj}(\bm{k}) = \sum_{j'} (\mathscr{A}_{\bm{k}})_{nj'} (\mathscr{S}_{\bm{k}}^{-1/2})_{j'j}.
\end{equation}
The Wannier orbitals can be shown to inherit the transformation properties of the trial orbitals under the symmetries of the Hamiltonian. In the Supplemental Material \cite{SuppMat} we demonstrate that the representations Eqs.~(\ref{eq:C6_wannier_rep},\ref{eq:U_wannier_rep},\ref{eq:A_wannier_rep}) may be enforced by the choice of trial orbitals
\begin{equation}
    g_{sj}(r,\phi)=e^{i(j-\frac{s}{2})\phi}e^{-(r-\mu)^{2}/2\xi^{2}}
\end{equation}
where the integer-valued term $j-s/2 = j\pm1/2$ has the interpretation of orbital angular momentum. In Fig.~\ref{fig:wannier_functions} we plot the resulting Wannier function densities corresponding to trial parameters with annular peak at $\mu=R$ and spread equal to the width of skyrmion wall, $\xi=R_0-R.$ This choice of the parameters $(\mu,\xi)$ yields a favorable ratio of the maximum and minimum values of $\det S_{\bm{k}}$ over the BZ, namely $<2$ for all parameters $(\alpha,J)$ considered, therefore indicating a smooth fit. Remarkably, the same Wannier functions result from instead inputting Gaussian trial functions $\mu=0$ with various $\xi,$ indicating that the annular features of the Wannier functions are a product of the Hamiltonian and not the trial functions. Further indication of our successful application of the projection method is provided by the Wannier functions $d_{s\bm{0}j}(\bm{r})$ decaying exponentially with distance from the skyrmion center, which we verified numerically over the range of several unit cells.

The Wannier functions Fig.~\ref{fig:wannier_functions}(a) are found to localize around the radius $R$ at which the Dirac mass $n_z({r})$ changes sign. Moreover, the states localize more tightly around $R$ as the Hund's coupling $J$ is increased. These features are consistent with the behavior of the previously studied single-skyrmion bound states \cite{2015_Hurst_Skyrmion_BS, KunalSkyrmion}. Where the lattice problem differs, however, is in the breaking of continuous rotational symmetry in the Wannier states at large $R\!\to\! R_0\!\sim\! a/2$ where the hybridization between the single-skyrmion bound states is largest due to proximity.

\subsubsection{Nearest-neighbor toy model}\label{NN-two-band-model}
When the skyrmion cutoff radius is not too large compared to the Hund's length scale, $R_0 J_\mathrm{eff}/\hbar v_F \lesssim 3,$ only the two lowest-lying bands are uniformly gapped from continuum of states at higher energies. Correspondingly, it is shown in Ref.~\cite{2015_Hurst_Skyrmion_BS} that only two electronic bound states accompany an isolated skyrmion for sufficiently small skyrmion radius. These observations motivate further investigation into the two-band tight-binding problem in particular. We note that Chern insulators in similar two-band lattice models with local orbitals 
having nonzero angular momentum have been discussed previously in the context of spin-orbit coupled ferromagnets \cite{cook2014,baidya2016}.

In Fig.~\ref{fig:wannier_functions}(b) we plot the two-band hopping parameters as a function of the skyrmion radius for fixed $J=3$ and $R_0/a=0.49$. As discussed above and further detailed in the Supplemental Material \cite{SuppMat}, constraints from symmetry ensure that the parameters $\varepsilon,t_{jj'},$ and $t^\text{n}_{jj'}$ characterizing the hopping Hamiltonian up to next-nearest neighbor are all real. Crucially, we find that the nearest-neighbor terms $t_{jj'}$ are more prominent than longer-range terms for a wide range of parameters $(J,R_0)$, thereby motivating a full analytical investigation of the two-band \textit{nearest-neighbor} toy model.

\begin{figure}[t]
	\includegraphics[width=0.44\textwidth]{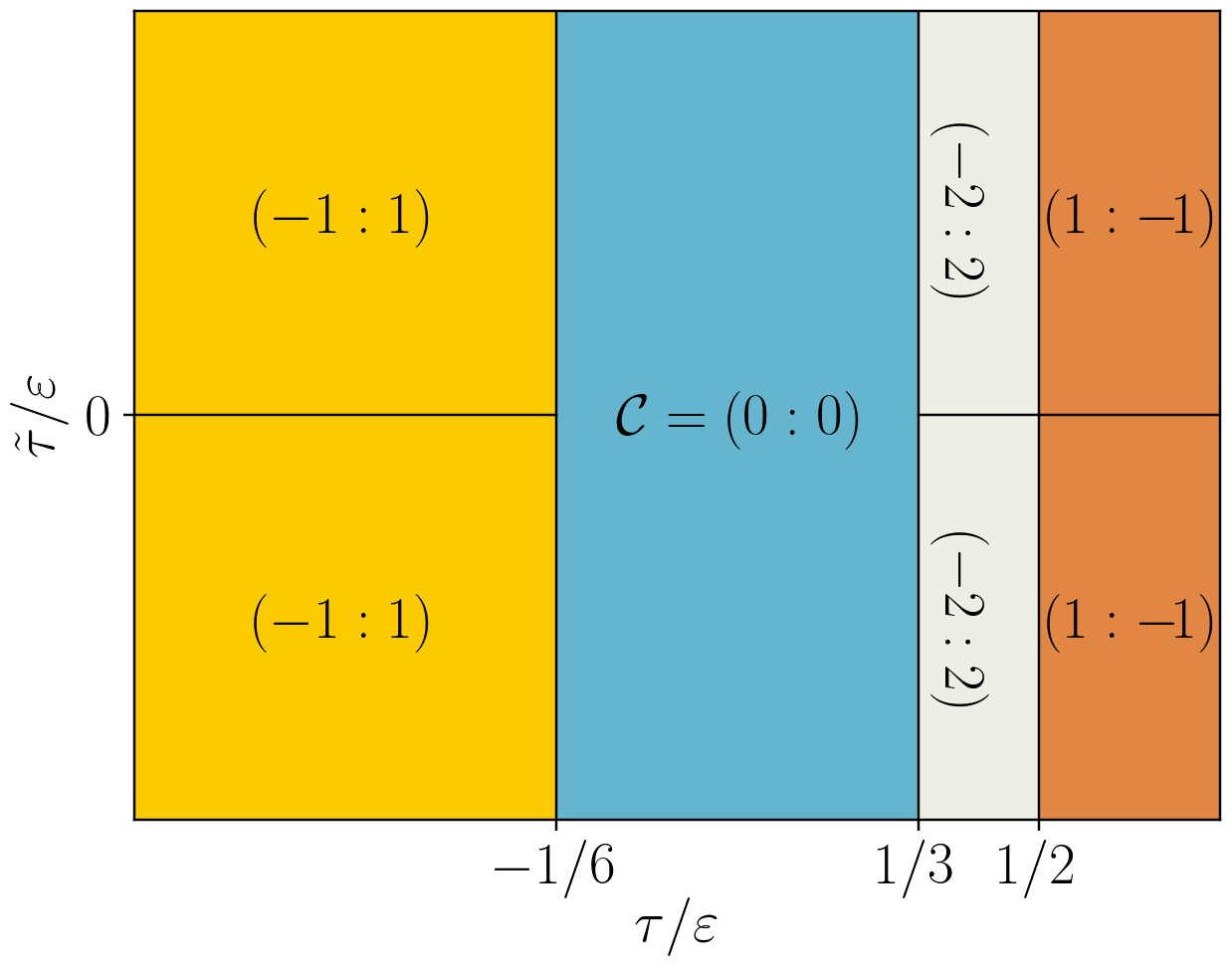}
	\caption{Topological phase diagram for the Bloch two-band nearest-neighbor toy model. We have assumed $\varepsilon>0,$ which is consistent with Fig.~\ref{fig:wannier_functions}(b), but the $\varepsilon<0$ phase diagram may be obtained by negating the Chern numbers. Both the $\mathcal{C}=(-1:1)$ and $(0:0)$ Chern sectors are realized in the continuum model, as exhibited in Fig.~\ref{fig:Bloch_chern_phase_diagram}.}
	\label{fig:2bandPD}
\end{figure}

In this case, the symmetry constraints Eqs.~(\ref{rotation},\ref{eq:energy-and-particle-hole},\ref{chiral_hoppings}) imply that the only free parameters are
\begin{equation}\begin{aligned}\label{eq:twoband_NN_freeparams}
    \varepsilon &\equiv \varepsilon_{-\!{1/2}} \\
    \tau &\equiv t_{{-\!{1/2}},{-\!{1/2}}} \\
    \tilde{\tau} &\equiv t_{{+\!{1/2}},{-\!{1/2}}}
\end{aligned}\end{equation}
and that they must all be real. The Hamiltonian Eq.~(\ref{eq:wannier_k_hamiltonian}) then takes the simple form $H_{\bm{k}}= h_{\bm{k}}\cdot\bm{\sigma},$ where
\begin{equation}
    h_{\bm{k}} \equiv \left(\tilde{\tau}\,\mathrm{Re}Q_{\bm{k}},\, -\tilde{\tau}\,\mathrm{Im}Q_{\bm{k}},\, -(\varepsilon+\tau P_{\bm{k}})\right), \label{eq:hk}
\end{equation}
and $P_{\bm{k}} \equiv s_{\bm{k}jj},$ $Q_{\bm{k}} \equiv s_{\bm{k}j\,j-1}.$ This gives rise to energies
\begin{equation}\label{eq:two-band-energies}
    E_{\bm{k}\pm} = \pm|h_{\bm{k}}| = \pm \sqrt{|\tilde{\tau} Q_{\bm{k}}|^2 + (\varepsilon + \tau P_{\bm{k}})^2}.
\end{equation}
Within this toy model, the Chern numbers of the bands $n=\pm$ are given by the number of times the unit vector $h_{\bm{k}}$ covers the unit sphere as the momentum is scanned through the BZ. Consequently, flipping the sign of $\tilde{\tau}$ has no bearing on the Chern number since it merely changes the helicity of the momentum space texture Eq.~(\ref{eq:hk}) and not its winding.
Due to continuity, the phase boundaries must occur at gap closures between the two bands, i.e. where $|h_{\bm{k}}|$ vanishes. Invoking Eq.~(\ref{eq:two-band-energies}), these gap closures are realized at the parameter values
\begin{equation}\label{eq:lines_locations}
    \tau/\varepsilon = -1/6,\, 1/3,\, 1/2,
\end{equation}
with the gap closures located at the $\Gamma$-point, three $M$-points, and two $K$-points respectively, as shown in the Supplemental Material \cite{SuppMat}. Special attention must be paid to the axis $\tilde{\tau}/\varepsilon = 0$ along which the gap is closed for $\tau/\varepsilon\geq 1/3$ and $\tau/\varepsilon \leq -1/6.$ Having identified the locations of all gap closures of the two-band toy model, we may compute the Chern numbers numerically. The results are summarized in Fig.~\ref{fig:2bandPD}.

Comparing the Chern number diagram to that which we derived from the continuum model, Fig.~\ref{fig:Bloch_chern_phase_diagram}, we observe that only the $\mathcal{C}=(-1:1)$ and $(0:0)$ Chern sectors are accessible by tuning the parameters $R_0,R,J$ of the continuum Hamiltonian. Moreover, we remark that the behavior at the boundary between these two Chern sectors is consistent between the continuum and toy models. As exhibited in Fig.~\ref{fig:wannier_functions}(b), which displays the hopping parameters obtained from the continuum model at $(J,R_0/a)=(3,0.49)$, truncating the tight-binding model at nearest-neighbor hopping causes only a small shift in the boundary between the two Chern sectors. Similar agreement is seen at other parameter values $(J,R_0)$. Finally, the gap closure at this transition is Dirac-like and occurs at the $\Gamma$ point in both the full tight-binding model and truncated toy model. We caution, however, that longer-range hoppings must be preserved in order to accurately fit to energies away from the $\Gamma$ point.
Finally, given the experimental capacity to simulate magnetic fluxes and the Hofstadter Hamiltonian using atoms in optical lattices \cite{Aidelsburger2013}, we note that it would be interesting to attempt to realize our family of tight-binding models in the ultracold atom setting.

\begin{figure*}
    \includegraphics[width=0.96\textwidth]{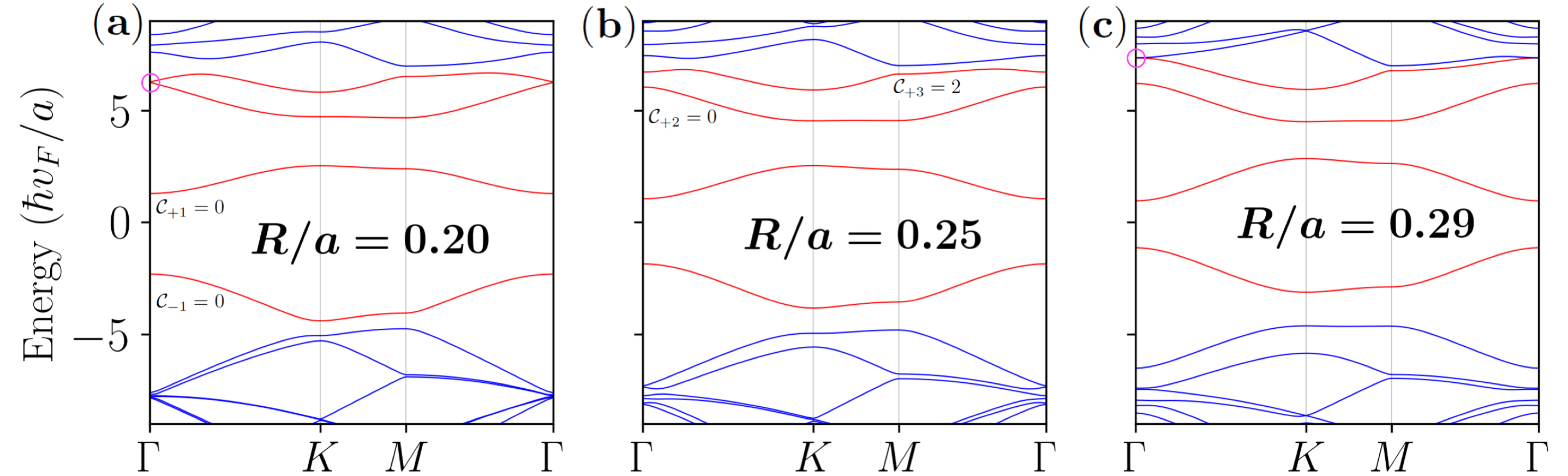}
    \caption{N\'eel skyrmion bands and density of states (DOS) at Hund's coupling $J=3$ and cutoff radius $R_0/a=0.3.$ Colored red are the bands whose Chern numbers are displayed in Fig.~\ref{fig:Neel_chern_phase_diagram}, namely the first band below half-filling and the three bands above it. \textbf{(a)} Whereas the bands carry Chern number $\mathcal{C}=(0:0,1,1)$ at the minimum value $R=R_0/2=0.15a$ of the core size, increasing $R$ leads to a gap closure (circled) between the $n=2,3$ bands. \textbf{(b)} These four bands enter the $\mathcal{C}=(0:0,0,2)$ Chern sector. The nontrivial $\mathcal{C}_{+3}=2$ Chern band hovers near the continuum before eventually closing a gap and descending toward zero energy. \textbf{(c)} While the N\'eel bands generically do not exhibit the particle-hole symmetry that they enjoyed in the Bloch case, we observe its reemergence in the limit $R\!\to\! R_0$ where both skyrmion types approach a domain wall droplet.}
    \label{fig:Neel_J=3_bands}
\end{figure*}

\section{N\'eel skyrmion lattice}

In contrast to the Bloch skyrmion case, the N\'eel in-plane divergence $\nabla^\mathrm{2D}\cdot\bm{n}(\bm{r})$ does not vanish identically but only in total when integrated over a single unit cell \cite{Nagaosa2010}. Consequently, there exists no gauge transformation which removes the in-plane magnetization component, thereby sacrificing both the particle-hole and chiral symmetries. The bands in the N\'eel case are therefore not symmetric across zero energy and do not have related Berry curvature.

Nonetheless, computation of the band eigenstates again reveals isolated low-energy bands with non-trivial Chern number. As in the Bloch case, tuning the skyrmion proximity, radial profile, and the effective Hund's coupling leads to a collection of phases distinguished by these Chern numbers. In Fig.~\ref{fig:Neel_J=3_bands} we plot the bands for fixed parameters $(J,R_0/a) = (3, 0.3)$ and for three values of the skyrmion size $R.$ Two of these values $R/a=0.2,0.29$ sit on the boundaries between topological sectors. In Fig.~\ref{fig:Neel_J=3_bands}(c) we observe the emergence of particle-hole symmetry. This is due to the Bloch and N\'eel skyrmions losing their distinction in the limit $R\to R_0,$ with both textures approaching a domain wall droplet with no in-plane magnetization.

\begin{figure}[b]
    \subfloat{\includegraphics[width=0.42\textwidth]{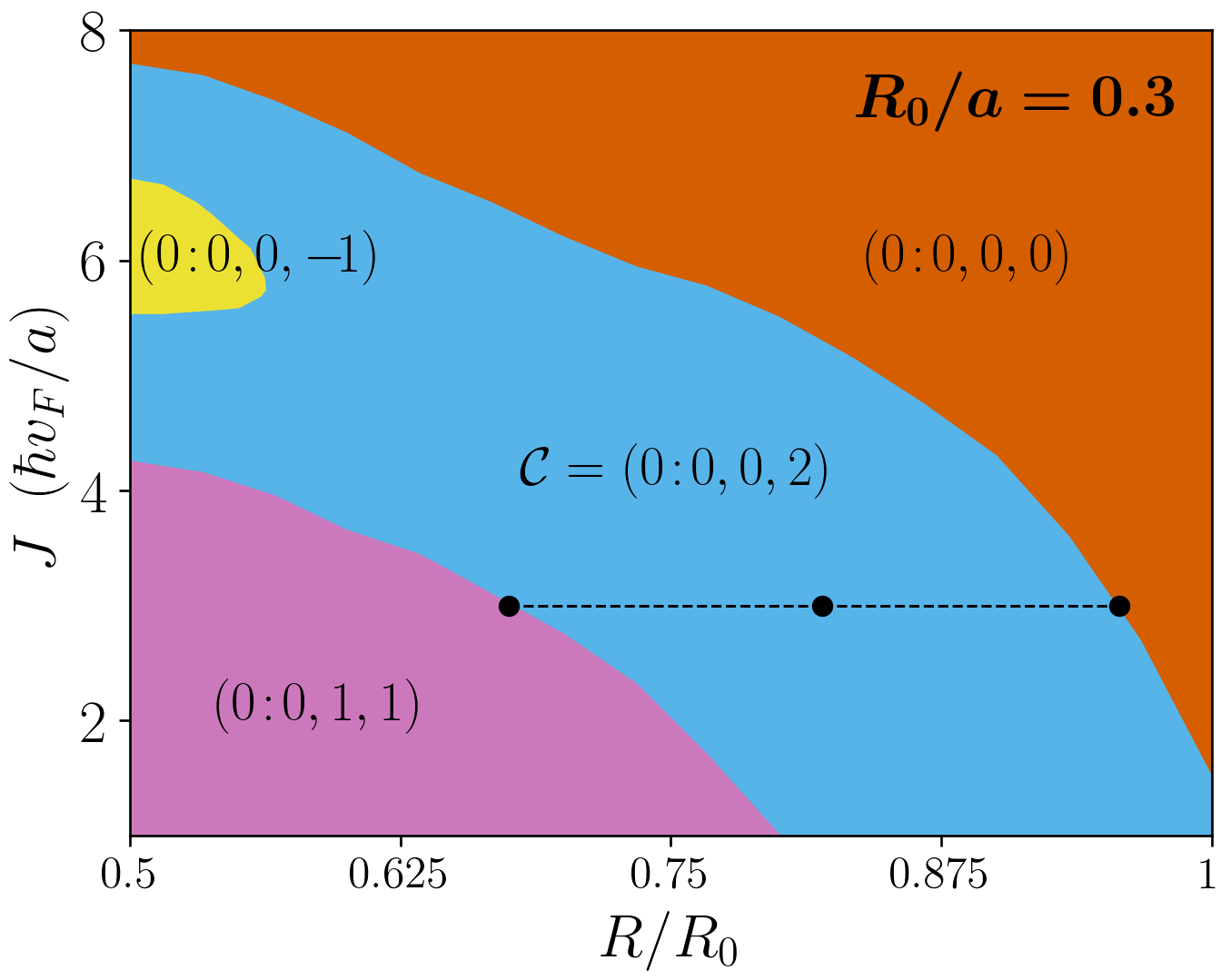}}
    \caption{$R_0/a=0.3.$ Chern number phase diagram in the N\'eel skyrmion case for a single band below half-filling and three bands above half-filling. The Chern numbers in each tuplet $\mathcal{C}$ are listed in order of increasing energy, with the colon marking half-filling. The marked points correspond to the bandstructure plots presented in Fig.~\ref{fig:Neel_J=3_bands}.}
    \label{fig:Neel_chern_phase_diagram}
\end{figure}

In Fig.~\ref{fig:Neel_chern_phase_diagram} we display the full topological phase diagram for fixed $R_0/a=0.3.$ Let us continue to label the bands by integers $n$ where $n<0$ ($n>0$) denote the bands below (above) half-filling. We find that the bands $n\in \{-1,1,2,3\}$ are consistently gapped from the remaining bands and therefore have well-defined Chern numbers. We denote them by $(\mathcal{C}_{-\!1} \!:\! \mathcal{C}_1, \mathcal{C}_2, \mathcal{C}_3)$ with the colon marking half-filling. The transition between the $\mathcal{C}=(0\!:\!0, 1, 1)$ and $(0\!:\!0,0,2)$ sectors upon increasing $R$ and $J$ occurs due to a gap closure at the $\Gamma$ point, Fig.~\ref{fig:Neel_J=3_bands}(a). Two further topological sectors are then accessible, including an island of $\mathcal{C}=(0\!:\!0, 0, -\!1)$ near the minimal value $R = R_0/2,$ as well as a robust Chern-trivial region which persist for large $J.$ We remark that the chosen value of $R_0/a=0.3$ showcases that Chern bands can be obtained in the N\'eel case for smaller cutoff radii $R_0$ and smaller skyrmion core sizes $R$ than in the Bloch case.

\section{Summary and Future Directions}
We have determined the band structure of TI helical Dirac surface states coupled to skyrmion crystal textures, revealing a strong dependence on the single-skyrmion radial profile $n_z$ which we studied by interpolating continuously between a sinsusoidal and domain wall limit. This stands in contrast to previous single-skyrmion \cite{2015_Hurst_Skyrmion_BS, Araki} and skyrmion lattice studies \cite{QAHE_graphene_skyrmions, paul2021topological} in which the textures were approximated by domain wall droplets. The question of tuning the radial profile, or of determining its most stable realization, is likely material-dependent and remains open to further investigation.

Significant qualitative differences between N\'eel and Bloch-type skyrmions were also elucidated. In the latter case, additional particle-hole and chiral symmetries led to constraints on the eigenstates and Berry curvature of the associated bands, producing a rich topological phase diagram upon varying the skyrmion separation, core size, and the spin-fermion coupling. In both cases we demonstrated the appearance of topologically non-trivial bands whose Chern numbers depend sensitively on the skyrmion radial profile. This intrinsic contribution to the Hall conductance contrasts with the topological Hall effect observed when nonrelativistic electrons couple to a skyrmion texture \cite{THE1}.

Going beyond the work described in this paper, we expect the coupling between skyrmion texture and fermions to be anisotropic \cite{TIdw}, with the coupling to $n_z$ being different from that to $n_x,n_y$. For Bloch skyrmions, this anisotropy has no impact since the in-plane component of the skyrmion texture can still be gauged away. It may, however, be interesting to investigate the impact of this anisotropy on the bands of N\'eel skyrmion crystals.

Using band theory techniques, we have constructed localized, symmetric Wannier orbitals for Dirac surface states coupled to a skyrmion crystal. To the best of our knowledge, such Wannier functions have not been extracted in previous work on this subject. For well-separated skyrmions, we have shown that the Wannier functions are `ring'-like states which resemble the previously studied single-skyrmion bound states. Within a two-band study, the truncation of the associated tight-binding model to nearest-neighbor hopping is found to capture the most relevant topological features.

In future studies it could prove fruitful to explore the effects of electron-electron interactions in such lattice models, which could support fractional Chern insulator phases \cite{FCI_Regnault} or other correlated states due to the presence of narrow Chern bands at large skyrmion core size and effective Hund's coupling strength. Furthermore, it would be interesting to explore how external magnetic fields tune the density and core size of skyrmions in the lattice, thereby allowing for systematic exploration of the topological phase diagrams discussed above. In this work, we assumed that a weak external magnetic field is sufficient to stabilize skyrmions \cite{panagopoulos2017,Chacon2018}. However, in materials where stronger external fields are needed to stabilize skyrmions, the Landau levels of the helical Dirac surface states could begin to play an important role in the skyrmion-skyrmion interaction as well as the nature of electronic states, leading to distinct topological features deserving of a separate investigation.

{\it Note added.} Recently, we came across a related paper \cite{paul2021topological} which examines the impact of spiral textures on TI Dirac surface states and the energetic stabilization of N\'eel skyrmions in intrinsically magnetic TIs. Our band theory calculations for N\'eel and Bloch skyrmion crystals, construction of Wannier orbitals, and tight-binding model results provide a complementary perspective on this topic.

\medskip
\medskip
\acknowledgements
T.P.-B. acknowledges useful discussions with Yafis Barlas, Kunal Tiwari, and William Coish, and funding from NSERC and FRQNT. S.D. and A.P. acknowledge funding from NSERC and useful discussions with Sopheak Sorn. S.D. acknowledges helpful discussions with Shubhayu Chatterjee, Stephen Gant, and Zachary Weinstein. S.D. is supported by the NSERC PGSD fellowship. H.L. thanks Richard Man-Wai Ling for his unfailing and enduring support.


\pagebreak
\widetext
\begin{center}
	\textbf{\large Supplemental Material: Magnetic skyrmion crystal at a topological insulator surface}
\end{center}
\setcounter{equation}{0}
\setcounter{figure}{0}
\setcounter{table}{0}
\setcounter{page}{1}
\makeatletter
\renewcommand{\theequation}{S\arabic{equation}}
\renewcommand{\thefigure}{S\arabic{figure}}
\renewcommand{\bibnumfmt}[1]{[S#1]}
\renewcommand{\citenumfont}[1]{S#1}

\setcounter{section}{0}
\renewcommand{\thesection}{S-\Roman{section}}

\section{Electron and Hole Vacua}

\label{sec:Bloch_and_Wannier} Discrete translational symmetry of
the Hamiltonian Eq.~(\ref{eq:continuum_Ham} of the main text) dictates that the energy
eigenstates will be labeled by band $n$ and crystal momentum $\bm{k}\in\mathrm{BZ}.$
Let $|\Omega\rangle$ denote the electron vacuum, i.e. the state which
is annihilated by all electron operators $c_{s\bm{G}}(\bm{k}).$ Then
the energy eigenstates are given by $\psi_{\bm{k}n}^{\dagger}|\Omega\rangle.$
Invoking Eq.~(\ref{eq:Bloch_expansion}), then the corresponding
Bloch wavefunction is 
\begin{equation}\label{eq:bloch_planewave}
\psi_{s\bm{k}n}(\bm{r})\equiv\langle\Omega|c_{s}(\bm{r})\psi_{\bm{k}n}^{\dagger}|\Omega\rangle=e^{i\bm{k}\cdot\bm{r}}u_{s\bm{k}\,n}(\bm{r}),
\end{equation}
where the cell-periodic part of the Bloch state is defined as 
\begin{equation}\label{eq:cellperiodic_bloch}
u_{s\bm{k}n}(\bm{r})\equiv\sum_{\bm{G}}u_{s\bm{G}n}(\bm{k})e^{i\bm{G}\cdot\bm{r}}.
\end{equation}
Appearing in later calculations is the hole vacuum $|\mho\rangle,$
the state annihilated by all of the operators $c_{s\bm{G}}^{\dagger}(\bm{k}).$
By again considering the expansion Eq.~(\ref{eq:Bloch_expansion}),
one may directly verify that $\langle\mho|\psi_{\bm{k}n}^{\dagger}c_{s}(\bm{r})|\mho\rangle=\psi_{s\bm{k}n}(\bm{r}).$
Likewise, for the Wannier states we have
\begin{equation}\label{eq:two_vacua_matrixelements}
\langle\Omega|c_{s}(\bm{r})d_{\bm{k}j}^{\dagger}|\Omega\rangle\equiv d_{s\bm{k}j}(\bm{r})=\langle\mho|d_{\bm{k}j}^{\dagger}c_{s}(\bm{r})|\mho\rangle
\end{equation}
due to their relation $d_{\bm{k}j}^{\dagger}=\sum_{n}\psi_{\bm{k}n}^{\dagger}\mathscr{U}_{nj}(\bm{k})$
to the Bloch states.

\medskip
\section{Six-fold rotational symmetry}

\subsection{Proof of symmetry in the continuum model}

\label{sec:C6_proof} Here we verify that the unitary $C_{6}$ defined
in Eq.~(\ref{eq:C6}) is a symmetry of the continuum Hamiltonian,
Eq.~(\ref{eq:continuum_Ham}). We remark that Eq.~(\ref{eq:C6})
may be equivalently represented in momentum space by 
\begin{equation}
C_{6}c_{G}^{\dagger}(\bm{k})C_{6}^{\dagger}=c_{C_{6}G}^{\dagger}(C_{6}\bm{k})e^{-i\frac{\pi}{3}\frac{\sigma_{z}}{2}}.
\end{equation}
Consequently, 
\begin{equation}
\begin{aligned}C_{6}HC_{6}^{\dagger} & =\sum_{\bm{k}}\sum_{\bm{GG}'}c_{C_{6}\bm{G}}^{\dagger}(C_{6}\bm{k})e^{-i\frac{\pi}{3}\frac{\sigma_{z}}{2}}(\mathscr{H}_{\bm{k}})_{\bm{G},\bm{G}'}e^{i\frac{\pi}{3}\frac{\sigma_{z}}{2}}c_{C_{6}\bm{G}'}(C_{6}\bm{k})\\
& =\sum_{\bm{k}}\sum_{\bm{GG}'}c_{G}^{\dagger}(\bm{k})\left[e^{-i\frac{\pi}{3}\frac{\sigma_{z}}{2}}(\mathscr{H}_{C_{6}^{-1}\bm{k}})_{C_{6}^{-1}\bm{G},C_{6}^{-1}\bm{G}'}e^{i\frac{\pi}{3}\frac{\sigma_{z}}{2}}\right]c_{\bm{G}'}(\bm{k}).
\end{aligned}
\end{equation}
Examining the Rashba term of Eq.~(\ref{eq:H_qGG}) we find by manipulating
Pauli matrices that 
\begin{equation}
\begin{aligned}e^{-i\frac{\pi}{3}\frac{\sigma_{z}}{2}}\left[\delta_{C_{6}^{-1}\bm{G},C_{6}^{-1}\bm{G}'}\left(C_{6}^{-1}\left(\bm{k}+\bm{G}\right)\times\bm{\sigma}\right)\cdot\hat{z}\right]e^{i\frac{\pi}{3}\frac{\sigma_{z}}{2}} & =\delta_{\bm{G},\bm{G}'}\left(C_{6}^{-1}\left(\bm{k}+\bm{G}\right)\times C_{6}^{-1}\bm{\sigma}\right)\cdot\hat{z}\\
& =\delta_{\bm{G},\bm{G}'}\left(\left(\bm{k}+\bm{G}\right)\times\bm{\sigma}\right)\cdot\hat{z}
\end{aligned}
\end{equation}
since $C_{6}$ is a rotation about $\hat{z}.$ For the Hund's coupling
we similarly have 
\begin{equation}
e^{-i\frac{\pi}{3}\frac{\sigma_{z}}{2}}\left[\bm{n}_{C_{6}^{-1}(\bm{G}-\bm{G}')}\cdot\bm{\sigma}\right]e^{i\frac{\pi}{3}\frac{\sigma_{z}}{2}}=\bm{n}_{C_{6}^{-1}(\bm{G}-\bm{G}')}\cdot C_{6}^{-1}\bm{\sigma}=\left(C_{6}\bm{n}_{C_{6}^{-1}(\bm{G}-\bm{G}')}\right)\cdot\bm{\sigma}
\end{equation}
which equals $\bm{n}_{\bm{G}-\bm{G}'}\cdot\bm{\sigma}$ since $C_{6}\bm{n}(C_{6}^{-1}\bm{r})=\bm{n}(\bm{r}).$
Therefore we have proved the matrix relation 
\begin{equation}
(\mathscr{H}_{\bm{k}})_{\bm{G},\bm{G}'}=e^{-i\frac{\pi}{3}\frac{\sigma_{z}}{2}}(\mathscr{H}_{C_{6}^{-1}\bm{k}})_{C_{6}^{-1}\bm{G},C_{6}^{-1}\bm{G}'}e^{i\frac{\pi}{3}\frac{\sigma_{z}}{2}},\qquad\text{i.e.}\ [C_{6},H]=0\label{eq:C6_matrix_constraint}
\end{equation}
whose immediate consequence is that the Bloch vectors $u_{sC_{6}\bm{G}\,n}(C_{6}\bm{k})$
and $u_{s\bm{G}\,n}(\bm{k})$ have degenerate eigenvalues and are
related by 
\begin{equation}
e^{i\frac{\pi}{3}\frac{s}{2}}u_{sC_{6}\bm{G}\,n}(C_{6}\bm{k})=e^{i\eta_{n}(\bm{k})}u_{s\bm{G}\,n}(\bm{k})
\end{equation}
for some function $\eta_{n}(\bm{k}).$ By Eqs.~(\ref{eq:bloch_planewave},\ref{eq:cellperiodic_bloch}), this likewise imposes the following gauge
constraint on the Bloch functions:
\begin{equation}
e^{i\frac{\pi}{3}\frac{s}{2}}\psi_{sC_{6}\bm{k}n}(C_{6}\bm{r})=e^{i\eta_{n}(\bm{k})}\psi_{s\bm{k}n}(\bm{r}).
\end{equation}

\subsection{Wannier function constraint}

\label{sec:C6_consequences} In this section we derive the constraint
that Eq.~(\ref{eq:C6_wannier_rep}), namely $d_{C_{6}\bm{R}j}^{\dagger}=e^{i\frac{\pi}{3}j}C_{6}d_{\bm{R}j}^{\dagger}C_{6}^{\dagger},$
places on the Wannier wavefunctions. Taking this operator equality
in $\langle\Omega|c_{s}(\bm{r})\dots|\Omega\rangle,$ we simply obtain
\begin{equation}\begin{aligned}\label{eq:C6_wfunc_constraint}
d_{sC_{6}\bm{R}j}(\bm{r}) & =e^{i\frac{\pi}{3}j}\langle\Omega|c_{s}(\bm{r})C_{6}d_{\bm{R}j}^{\dagger}C_{6}^{\dagger}|\Omega\rangle\\
& =e^{i\frac{\pi}{3}j}\langle\Omega|\left(C_{6}e^{-i\frac{\pi}{3}\frac{s}{2}}c_{s}(C_{6}^{-1}\bm{r})\right)d_{\bm{R}j}^{\dagger}C_{6}^{\dagger}|\Omega\rangle\\
& =e^{i\frac{\pi}{3}(j-\frac{s}{2})}d_{s\bm{R}j}(C_{6}^{-1}\bm{r})
\end{aligned}\end{equation}
from the definition of $C_{6}$ on the continuum operators, Eq.~(\ref{eq:C6}).

If the trial Wannier functions $g_{sj}(\bm{r}),$ which are centered
in the home unit cell $\bm{R}=\bm{0},$ satisfy this same relation
\begin{equation}\label{eq:g_constraint_C6}
g_{sj}(\bm{r})=e^{i\frac{\pi}{3}(j-\frac{s}{2})}g_{sj}(C_{6}^{-1}\bm{r})
\end{equation}
then we obtain the following overlap between $g$ and the Bloch states:
\begin{equation}
\begin{aligned}(A_{\bm{k}})_{nj} & =\sum_{s=\pm1}\int_{\R^{2}}d^{2}\bm{r}\ \psi_{s\bm{k}n}^{*}(\bm{r})g_{sj}(\bm{r})\\
& =\sum_{s=\pm1}\int_{\R^{2}}d^{2}\bm{r}\ e^{i\eta_{n}(\bm{k})}\left(\psi_{C_{6}\bm{k}}^{*}(C_{6}\bm{r})e^{-i\frac{\pi}{3}\frac{s}{2}}\right)g_{j}(\bm{r})\\
& =e^{i\eta_{n}(\bm{k})}\left(\sum_{s=\pm1}\int_{\R^{2}}d^{2}\bm{r}\ \psi_{C_{6}\bm{k}n}^{*}(\bm{r})g_{j}(\bm{r})\right)e^{-i\frac{\pi}{3}j}\\
& =e^{i\eta_{n}(\bm{k})}(A_{C_{6}\bm{k}})_{nj}e^{-i\frac{\pi}{3}j}.
\end{aligned}    
\end{equation}
Hence, $\mathscr{U}(\bm{k})\equiv A_{\bm{k}}(A_{\bm{k}}^{\dagger}A_{\bm{k}})^{-1/2}\implies\mathscr{U}_{nj}(\bm{k})=e^{i\eta_{n}(\bm{k})}\mathscr{U}_{nj}(C_{6}\bm{k})e^{-i\frac{\pi}{3}j}$
where $\mathscr{U}(\bm{k})$ is the unitary taking us from Bloch to
Wannier states via Eq.~(\ref{eq:wannier_to_bloch_unitary}), namely $d_{s\bm{k}\,j}(\bm{r})=\sum_{n}\psi_{s\bm{k}\,n}(\bm{r})\mathscr{U}_{nj}(\bm{k}).$
We therefore find that
\begin{equation}\begin{aligned}
d_{sC_{6}\bm{k}j}(\bm{r}) & =\sum_{n}\psi_{sC_{6}\bm{k}\,n}(\bm{r})\mathscr{U}_{nj}(C_{6}\bm{k})\\
& =\sum_{n}\left(e^{i\eta_{n}(\bm{k})}\psi_{s\bm{k}n}(C_{6}^{-1}\bm{r})e^{-i\frac{\pi}{3}\frac{s}{2}}\right)\mathscr{U}_{nj}(C_{6}\bm{k})\\
& =e^{-i\frac{\pi}{3}\frac{s}{2}}\sum_{n}\psi_{s\bm{k}n}(C_{6}^{-1}\bm{r})e^{i\frac{\pi}{3}j}\mathscr{U}_{nj}(\bm{k})\\
& =e^{i\frac{\pi}{3}(j-\frac{s}{2})}d_{s\bm{k}j}(C_{6}^{-1}\bm{r})
\end{aligned}\end{equation}
which is precisely the Fourier transform of Eq.~(\ref{eq:C6_wfunc_constraint}).
In summary, we have demonstrated that imposing the $C_{6}$ condition
on the origin-centered Wannier trial functions $g$ enforces that
same condition on the Wannier functions constructed from the projection
method. Similar logic holds for the particle-hole and chiral symmetries discussed later, though some care must be taken because of the exchange of electrons and holes.

\medskip
\section{Particle-hole symmetry}

\subsection{Proof of symmetry in the continuum model}

\label{sec:PH_proof} Here we verify that the unitary $U$ defined
by Eq.~(\ref{eq:PH}) is in fact a symmetry of the continuum Hamiltonian
Eq.~(\ref{eq:continuum_Ham}). The conjugated Hamiltonian is
\begin{equation}
\begin{aligned}UHU^{\dagger} & =\sum_{\bm{k}}\left(\sigma_{y}c(\bm{k})\right)\,\mathscr{H}_{\bm{k}}\,\left(c^{\dagger}(\bm{k})\sigma_{y}\right)\\
& =\sum_{\bm{k}}c(\bm{k})\sigma_{y}^{T}\,\mathscr{H}_{\bm{k}}\,\sigma_{y}^{T}c^{\dagger}(\bm{k})\\
& =\sum_{\bm{k}}c^{\dagger}(\bm{k})\left[-\sigma_{y}\,(\mathscr{H}_{\bm{k}})^{*}\,\sigma_{y}\right]c(\bm{k})
\end{aligned}
\end{equation}
plus a constant which vanishes when the single-particle operators in $UHU^{\dagger}$
and $H$ are equal. In a manner similar to Eq.~(\ref{eq:C6_matrix_constraint}),
one can show that the property $\mathbbm{n}(\bm{r})=\mathbbm{n}(-\bm{r})$
is sufficient to guarantee
\begin{equation}
-\sigma_{y}\,(\mathscr{H}_{\bm{k}})^{*}\,\sigma_{y}=\mathscr{H}_{\bm{k}}.
\end{equation}
This proves that $U$ is a symmetry, and moreover implies that the energies
at each $\bm{k}$ are symmetric about zero $\epsilon_{-n}(\bm{k})=-\epsilon_{n}(\bm{k})$
with corresponding Bloch vectors related by
\begin{equation}
u_{-n}^{*}(\bm{k}) = e^{i\beta_{n}(\bm{k})} \sigma_{y}u_{n}(\bm{k})
\end{equation}
for some function $\beta_{n}(\bm{k}).$ The energy eigenfunctions
can then be computed via Eqs.~(\ref{eq:bloch_planewave},\ref{eq:cellperiodic_bloch}), yielding a corresponding relationship
\begin{equation}
\psi_{\bm{k}\,-\!n}^{*}(-\bm{r}) = e^{i\beta_{n}(\bm{k})} \sigma_{y}\psi_{\bm{k}n}(\bm{r}).\label{eq:U_bloch_proportionality}
\end{equation}

\subsection{Consequences for the Wannier functions}

\label{sec:PH_consequences}

In this section we detail the constraint that Eq.~(\ref{eq:U_wannier_rep}),
namely $e^{i\pi j}d_{\bm{k}\,-\!j}^{\dagger}=Ud_{\bm{k}j}U^{\dagger},$
places on the Wannier wavefunctions. To compare the wavefunctions,
we sandwich both sides in $\langle\Omega|c_{s}(\bm{r})\dots|\Omega\rangle.$
Since $U^{\dagger}$ exchanges particles and holes then, up to a phase, it maps
the electron vacuum $|\Omega\rangle$ to the hole vacuum $|\mho\rangle.$
Letting $\sigma_y$ act on the spin index, then
\begin{equation}
\begin{aligned}\label{eq:U_wannierfunc_symm_k}
e^{i\pi j}\langle\Omega|c(\bm{r})d_{\bm{k}\,-\!j}^{\dagger}|\Omega\rangle & =\langle\Omega|c(\bm{r})Ud_{\bm{k}j}U^{\dagger}|\Omega\rangle \\
e^{i\pi j}d_{\bm{k}\,-\!j}(\bm{r}) & =\langle\Omega|\left(\sigma_{y}Uc^{\dagger}(-\bm{r})\right)d_{\bm{k}j}U^{\dagger}|\Omega\rangle \\
& =\sigma_{y}\left(\langle\mho|d_{\bm{k}j}^{\dagger}c(-\bm{r})|\mho\rangle\right)^{*} \\
& =\sigma_{y}d_{\bm{k}j}^{*}(-\bm{r})
\end{aligned}
\end{equation}
by Eq.~(\ref{eq:two_vacua_matrixelements}), where we moreover invoked the real space representation of Eq.~(\ref{eq:PH}),
namely $c(\bm{r})U=\sigma_{y}Uc^{\dagger}(-\bm{r}).$ By Fourier transformation, this imposes the
following constraint on the Wannier functions:
\begin{equation}
e^{i\pi j}d_{\bm{R}\,-\!j}(\bm{r})=\sigma_{y}d_{-\!\bm{R}\,j}^{*}(-\bm{r}). \label{eq:U_wannierfunc_symm_constraint}
\end{equation}

Following the strategy of Sec.~\ref{sec:C6_consequences}, we demand
that the origin-centered trial orbital $g_{sj}(\bm{r})$ satisfies
this symmetry constraint, namely
\begin{equation}\label{eq:g_constraint_U}
e^{i\pi j}g_{-\!j}(\bm{r})=\sigma_{y}g_{j}^{*}(-\bm{r}).    
\end{equation}
We then verify, as a result, that the constructed Wannier functions satisfy the constraint as well. The overlaps are given by
\begin{equation}\begin{aligned}
(A_{\bm{k}})_{nj} & =\int_{\R^{2}}d^{2}\bm{r}\ \psi_{\bm{k}n}^{*}(\bm{r})g_{j}(\bm{r})\\
& =\int_{\R^{2}}d^{2}\bm{r}\ \left(e^{i\beta_{n}(\bm{k})}\psi_{\bm{k}\,-\!n}(-\bm{r})\sigma_{y}\right)\left(e^{i\pi j}\sigma_{y}g_{-j}^{*}(-\bm{r})\right)\\
& =e^{i\beta_{n}(\bm{k})}\left[\int_{\R^{2}}d^{2}\bm{r}\ \psi_{\bm{k}\,-\!n}^{*}(-\bm{r})g_{-j}(-\bm{r})\right]^{*}e^{i\pi j}\\
& =e^{i\beta_{n}(\bm{k})}(A_{\bm{k}})_{-\!n,-\!j}^{*}e^{i\pi j}
\end{aligned}\end{equation}
so that $\mathscr{U}_{nj}(\bm{k})=e^{i\beta_{n}(\bm{k})}\mathscr{U}_{-\!n,-\!j}^{*}(\bm{k})e^{i\pi j}.$
Then Eq.~(\ref{eq:wannier_to_bloch_unitary}) and the Bloch state proportionality Eq.~(\ref{eq:U_bloch_proportionality}) imply that
\begin{equation}\begin{aligned}\label{eq:U_wannier_verification}
e^{i\pi j}d_{\bm{k}\,-\!j}(\bm{r}) & =e^{i\pi j}\sum_{n}\psi_{\bm{k}n}(\bm{r})\mathscr{U}_{n,-\!j}(\bm{k})\\
& =e^{i\pi j}\sum_{n}\psi_{\bm{k}n}(\bm{r})\left(e^{i\beta_{n}(\bm{k})}\mathscr{U}_{-\!n,j}^{*}(\bm{k})e^{-i\pi j}\right)\\
& =\sum_{n}\left(e^{i\beta_{n}(\bm{k})}\psi_{\bm{k}n}(\bm{r})\right)\mathscr{U}_{-\!n,j}^{*}(\bm{k})\\
& =\sum_{n}\left(\sigma_{y}\psi_{\bm{k}\,-\!n}^{*}(-\bm{r})\right)\mathscr{U}_{-\!n,j}^{*}(\bm{k})\\
& =\sigma_{y}d_{\bm{k}j}^{*}(-\bm{r})
\end{aligned}\end{equation}
which precisely matches Eq. \ref{eq:U_wannierfunc_symm_k}.

\medskip
\section{Chiral symmetry}

\label{sec:chiral_proof}

\subsection{Proof of symmetry in the continuum model}

Here we verify that the anti-unitary operator defined in Eq.~(\ref{eq:chiral})
is a symmetry of the continuum Hamiltonian, Eq.~(\ref{eq:continuum_Ham}).
Let $\mathscr{M}_{y}$ be a matrix which acts on the reciprocal vector
label, mapping $\bm{G}\to M_{y}\bm{G}.$ Employing this notation,
we compute 
\begin{equation}
\begin{aligned}AHA^{\dagger} & =\sum_{\bm{k}}\left(\mathscr{M}_{y}\sigma_{x}c(M_{y}\bm{k})\right)\mathscr{H}_{\bm{k}}^{*}\left(c^{\dagger}(M_{y}\bm{k})\mathscr{M}_{y}\sigma_{x}\right)\\
& =-\sum_{\bm{k}}\left(c^{\dagger}(M_{y}\bm{k})\mathscr{M}_{y}\sigma_{x}\right)(\mathscr{H}_{\bm{k}}^{*})^{T}\left(\mathscr{M}_{y}\sigma_{x}c(M_{y}\bm{k})\right)\\
& =\sum_{\bm{k}}c^{\dagger}(\bm{k})\left[-(\sigma_{x}\mathscr{M}_{y})\mathscr{H}_{M_{y}\bm{k}}(\sigma_{x}\mathscr{M}_{y})\right]c(\bm{k}).
\end{aligned}
\end{equation}
In a manner similar to Eq.~(\ref{eq:C6_matrix_constraint}), the
property $\mathbbm{n}(\bm{r})=\mathbbm{n}(M_{y}\bm{r})$ is sufficient
to guarantee that 
\begin{equation}
\mathscr{H}_{\bm{k}}=-(\sigma_{x}\mathscr{M}_{y})\mathscr{H}_{M_{y}\bm{k}}(\sigma_{x}\mathscr{M}_{y})
\end{equation}
so that $[A,H]=0.$ This relation moreover implies that the Bloch
vectors between the bands at mirror-related momenta must have opposite
energies 
\begin{equation}
-\epsilon_{n}(\bm{k})=\epsilon_{-n}(M_{y}\bm{k})
\end{equation}
and be related by 
\begin{equation}
e^{i\gamma_{n}(\bm{k})}u_{n}(\bm{k})=(\sigma_{x}\mathscr{M}_{y})u_{-n}(M_{y}\bm{k})\label{eq:My_proportionality}
\end{equation}
for some phase $\gamma_{n}(\bm{k}).$ By Eqs.~(\ref{eq:bloch_planewave},\ref{eq:cellperiodic_bloch})
we correspondingly find that the Bloch functions are related by 
\begin{equation}
e^{i\gamma_{n}(\bm{k})}\psi_{\bm{k}n}(\bm{r})=\sigma_{x}\psi_{M_{y}\bm{k}\,-\!n}(M_{y}\bm{r}).
\end{equation}

\subsection{Consequences for Berry Curvature and Chern numbers}

The chiral symmetry is distinct from $C_{6}$ and $U$ in that it
provides a useful constraint on the Berry curvatures and Chern numbers.
Let $\epsilon_{\mu\nu}$ denote the Levi-Civita symbol in 2D. We adopt
the convention where 
\begin{equation}
\mathcal{C}_{n}=\frac{1}{2\pi i}\int_{BZ}d^{2}\bm{k}\ F^{(n)}(\bm{k}),\qquad F^{(n)}(\bm{k})=\epsilon_{\mu\nu}\frac{\partial}{\partial\bm{k}^{\mu}}A_{\nu}^{(n)}(\bm{k})
\end{equation}
with the Berry connection given by 
\begin{equation}
\begin{aligned}A_{\mu}^{(n)}(\bm{k}) & =u_{n}^{\dagger}(\bm{k})\cdot\frac{\partial}{\partial\bm{k}^{\mu}}u_{n}(\bm{k})\\
& =u_{-n}^{\dagger}(M_{y}\bm{k})\left(\sigma_{x}\mathscr{M}_{y}\right)^{\dagger}\cdot\left(\sigma_{x}\mathscr{M}_{y}\right)\frac{\partial}{\partial\bm{k}^{\mu}}\left(u_{-n}(M_{y}\bm{k})\right)\\
& =u_{-n}^{\dagger}(M_{y}\bm{k})\cdot\left(M_{y}\right)_{\mu\alpha}\frac{\partial u_{-n}}{\partial\bm{k}^{\alpha}}\Big\vert_{M_{y}\bm{k}}.
\end{aligned}
\end{equation}
Since the Berry curvature is gauge-invariant then we may conveniently
take the phase in Eq.~(\ref{eq:My_proportionality}) to be unity
for the sake of comparing the Berry curvatures. This gives 
\begin{equation}
\begin{aligned}F^{(n)}(\bm{k}) & =\epsilon_{\mu\nu}\frac{\partial}{\partial\bm{k}^{\mu}}\left(u_{-n}^{\dagger}(M_{y}\bm{k})\cdot\left(M_{y}\right)_{\nu\alpha}\frac{\partial u_{-n}}{\partial\bm{k}^{\alpha}}\Big\vert_{M_{y}\bm{k}}\right)\\
& =\epsilon_{\mu\nu}\left(\left(M_{y}\right)_{\mu\beta}\frac{\partial}{\partial\bm{k}^{\beta}}\left(u_{-n}^{\dagger}\cdot\left(M_{y}\right)_{\nu\alpha}\frac{\partial u_{-n}}{\partial\bm{k}^{\alpha}}\right)\right)\Big\vert_{M_{y}\bm{k}}
\end{aligned}
\end{equation}
where $\epsilon_{\mu\nu}\left(M_{y}\right)_{\mu\beta}\left(M_{y}\right)_{\nu\alpha}=-\epsilon_{\beta\alpha}$
implies that 
\begin{equation}
\begin{aligned}F^{(n)}(\bm{k}) & =-\epsilon_{\beta\alpha}\left(\frac{\partial}{\partial\bm{k}^{\beta}}\left(v_{-n}^{\dagger}\cdot\frac{\partial v_{-n}}{\partial\bm{k}^{\alpha}}\right)\right)\Big\vert_{M_{y}\bm{k}}\\
& =-\epsilon_{\beta\alpha}\left(\frac{\partial}{\partial\bm{k}^{\beta}}A_{\alpha}^{(-n)}\right)\Big\vert_{M_{y}\bm{k}}\\
& =-F^{(-n)}(M_{y}\bm{k})
\end{aligned}
\end{equation}
as claimed in Eq.~(\ref{eq:Berry_constraint}). Integrating both
sides over the Brillouin zone, we immediately deduce that the Chern
numbers in particle-hole-related bands are opposite, 
\begin{equation}
\mathcal{C}_{-n}=-\mathcal{C}_{n}.
\end{equation}

\subsection{Consequences for the Wannier functions}

\label{sec:chiral_consequences} In this section we derive the constraint that Eq.~(\ref{eq:A_wannier_rep}), namely $Ad_{\bm{k}\,j}A^{\dagger}=d_{M_{y}\bm{k}\,-j}^{\dagger},$
places on the Wannier functions. As with the particle-hole operator $U$ above, the chiral operator $A$ exchanges the electron vacuum $|\Omega\rangle$ with the hole vacuum $|\mho\rangle$ up to an inconsequential phase. Following the route of Eq.~(\ref{eq:U_wannierfunc_symm_k}) and using the anti-unitarity of $A,$ then
\begin{equation}
d_{\bm{k},-\!j}(\bm{r})=\sigma_{x}d_{M_{y}\bm{k}\,j}(M_{y}\bm{r})    
\end{equation}
whose derivation requires the real-space representation of Eq.~(\ref{eq:chiral}),
namely $Ac(\bm{r})A^{\dagger}=c^{\dagger}(M_{y}\bm{r})\sigma_{x}.$
By Fourier transform, we find for the Wannier functions:
\begin{equation}\label{eq:A_wannfunc_constraint}
d_{\bm{R},-\!j}(\bm{r})=\sigma_{x}d_{M_{y}\bm{R}\,j}(M_{y}\bm{r}).    
\end{equation}

As in Sec.~(\ref{sec:C6_consequences}, \ref{sec:PH_consequences}) we proceed by imposing this constraint
on the trial Wannier functions,
\begin{equation}\label{eq:g_constraint_A}
g_{-j}(\bm{r})=\sigma_{x}g_{j}(M_{y}\bm{r})
\end{equation}
so as to ensure that Eq.~(\ref{eq:A_wannfunc_constraint}) is satisfied by the
constructed Wannier functions. Indeed, we find that the overlaps satisfy
$(A_{\bm{k}})_{nj}=e^{i\gamma_{n}(\bm{k})}(A_{M_{y}\bm{k}})_{-\!n\,-\!j},$
and likewise
\begin{equation}
\mathscr{U}_{nj}(\bm{k})=e^{i\gamma_{n}(\bm{k})}\mathscr{U}_{-\!n\,-\!j}(M_{y}\bm{k}).
\end{equation}
Proceeding as in Eq.~(\ref{eq:U_wannier_verification}), this is easily shown to enforce Eq.~(\ref{eq:A_wannfunc_constraint}).

\subsection{Selection of the trial functions for Wannier projection}
\label{sec:trial_functions}
We now detail an ansatz class of trial functions that obey the constraints Eqs.~(\ref{eq:g_constraint_C6}, \ref{eq:g_constraint_U}, \ref{eq:g_constraint_A}) derived above. As discussed in Sec.~\ref{subsc:single_skyrmion}, the single-skyrmion bound states are ring-like with angular dependence $e^{i\phi(j-\frac{s}{2})},$ therefore motivating us to take 
\begin{equation}
g_{sj}(r,\phi) = e^{i\phi(j-\frac{s}{2})} e^{-(r-\mu)^2/2\xi^2}
\end{equation}
where $(\mu,\xi)$ are free parameters. It is simple to verify that the required constraints are satisfied. In Sec.~\ref{sec:Wannier} we specify the values taken for $(\mu,\xi)$ and the Wannier functions that result.

\medskip
\section{Additional details of the tight binding model}
\subsection{Proof of equation (\ref{rotation-transform})}
\label{proof-of-rotation}
The $C_6$ representation Eq.~(\ref{eq:C6_wannier_rep}) allows us to relate the hopping amplitudes on links related by a $\pi/3$ rotation. From $C_6 d_{\bm{R}j}C_6^\dagger = e^{i\frac{\pi}{3}j} d_{C_6\bm{R}j},$ we have $C_6|\bm{R}j\rangle = e^{-i\frac{\pi}{3}j}|C_6\bm{R}\,j\rangle.$ Therefore,
\begin{equation}\begin{aligned}
t_{C_6\bm{\delta}\, j j'} & = \langle C_6\bm{\delta} \ j \rvert H \lvert \bm{0} j' \rangle \\
& = e^{-i\frac{\pi}{3}j} \langle \bm{\delta}  j  \rvert C_6^\dagger\ H\ C_6 \lvert \bm{0} j' \rangle e^{i\frac{\pi}{3}j' } \\
& = e^{-i\frac{\pi}{3} (j\!-\!j')} \langle \bm{\delta}  j \rvert  H \lvert \bm{0} j' \rangle \\
& = e^{-i\frac{\pi}{3} (j\!-\!j')} t_{\bm{\delta} j j'}.
\end{aligned}\end{equation}

\subsection{Definition and properties of $s_{k j j'}$}
\label{properties-of-s}
The function $s_{\bm{k} j j'}$ is the sum of phase factors related to nearest-neighbor hopping processes. It is defined as
\begin{equation}\begin{aligned}
s_{\bm{k} j j'} & \equiv e^{-i\bm{k}\cdot\bm{a}_1} + e^{-i\bm{k}\cdot\bm{a}_2}e^{-i\frac{\pi}{3}(j\! -\! j')} + e^{-i\bm{k}\cdot\bm{a}_3}e^{-i\frac{2\pi}{3}(j\! -\! j')} + \\ 
&\qquad e^{i\bm{k}\cdot\bm{a}_1}e^{-i \pi (j\! -\! j')} +  e^{i\bm{k}\cdot\bm{a}_2}e^{-i\frac{4\pi}{3}(j\! -\! j')} + e^{i\bm{k}\cdot\bm{a}_3}e^{-i\frac{5\pi}{3}(j\! -\! j')}. 
\end{aligned}\end{equation}
By direct substitution, we can verify the following properties:
\begin{eqnarray}
s_{\bm{k} j j'} & = & s_{\bm{k} -\!j' -\!j} \label{s-first}\\
s_{\bm{k} j j'} & = & s_{M_y\bm{k} -j\! -\!j'} \label{s-second}\\
s_{\bm{k} j' j} & = & \left( s_{\bm{k} j j'} \right)^* e^{-i\pi(j\!-\!j')}
\end{eqnarray}
Note that $s_{\bm{k} j\, j+m}$ depends only on $m$. It is straightforward to verify that
\begin{equation}
s_{\bm{k} j\, j+m}  =  2e^{i\frac{\pi m}{2}}\cos(\bm{k}\cdot\bm{a}_1+{\scriptstyle\frac{\pi m}{2}}) + 2e^{i\frac{5\pi m}{6}}\cos(\bm{k}\cdot \bm{a}_2 + {\scriptstyle\frac{\pi m}{2}}) + 2e^{i\frac{7\pi m}{6}} \cos ( \bm{k}\cdot \bm{a}_3 + {\scriptstyle\frac{\pi m}{2}}).
\end{equation}
For the special cases $m = 0,-1,-2,-3$ we have
\begin{equation}\begin{aligned}
s_{\bm{k} j\, j} & = 2\cos(\bm{k}\cdot\bm{a}_1) + 2\cos(\bm{k}\cdot \bm{a}_2) + 2\cos(\bm{k}\cdot \bm{a}_3) \\
s_{\bm{k} j\, j\!-\!1} & = 2e^{-i\frac{\pi}{2}}\sin(\bm{k}\cdot\bm{a}_1) + 2e^{-i\frac{5\pi}{6}}\sin(\bm{k}\cdot \bm{a}_2) + 2e^{-i\frac{7\pi}{6}} \sin(\bm{k}\cdot \bm{a}_3)  \\
s_{\bm{k} j\, j\!-\!2} & = 2\cos(\bm{k}\cdot\bm{a}_1) + 2e^{-i\frac{2\pi}{3}}\cos(\bm{k}\cdot \bm{a}_2) + 2e^{-i\frac{4\pi}{3}}\cos(\bm{k}\cdot \bm{a}_3)  \\
s_{\bm{k} j\, j\!-\!3} & = 2e^{-i\frac{\pi}{2}}\sin(\bm{k}\cdot\bm{a}_1) + 2e^{-i\frac{3\pi}{2}}\sin(\bm{k}\cdot \bm{a}_2) + 2e^{-i\frac{5\pi}{2}}\sin(\bm{k}\cdot \bm{a}_3).
\end{aligned}\end{equation}

\subsection{Proof of equation (\ref{eq:energy-and-particle-hole})}

\label{proof-of-particle-hole}Recalling from Eq.~(\ref{eq:wannier_k_hamiltonian}) that $H_{\bm{k}jj'}=\delta_{jj'}\varepsilon_{j}+s_{\bm{k}jj'}t_{jj'}$ at nearest-neighbor range, then
\begin{equation}\begin{aligned}
H_{\bm{k}-\!j'-\!j} & = \delta_{-\!j'-\!j}\,\varepsilon_{-\!j'}+s_{\bm{k}-\!j'-\!j}t_{-\!j'-\!j} \\
& = \delta_{jj'}\,\varepsilon_{-\!j}+s_{\bm{k}jj'}t_{-\!j'-\!j}
\end{aligned}\end{equation}
where we invoked Eq.~(\ref{s-first}). Therefore $H_{\bm{k}-\!j'-\!j}\,e^{i\pi(j'\!-\!j)}=-H_{\bm{k}jj'}$
if and only if $\varepsilon_{-j}=-\varepsilon_{j}$ and $t_{-\!j'\,-\!j}=-e^{i\pi(j\!-\!j')}t_{jj'}.$ This verifies the contraints on the tight-binding parameters due to particle-hole symmetry.

\subsection{Proof of equation (\ref{chiral_hoppings})}

\label{proof-of-chiral}We similarly compute
\begin{equation}\begin{aligned}
H_{M_{y}\bm{k}-\!j-\!j'} & = \delta_{-\!j-\!j'}\,\varepsilon_{-\!j}+s_{M_{y}\bm{k}-\!j-\!j'}t_{-\!j-\!j'}\\
& = \delta_{jj'}\,\varepsilon_{-\!j}+s_{\bm{k}jj'}t_{-\!j-\!j'} 
\end{aligned}\end{equation}
where we have used equation (\ref{s-second}). Therefore $H_{M_{y}\bm{k}-\!j-\!j'}=-H_{\bm{k}jj'}$
if and only if $\varepsilon_{-j}=-\varepsilon_{j}$ and $t_{-\!j\,-\!j'}=-t_{jj'}.$ This verifies the contraints on the tight-binding parameters due to the chiral symmetry.

\subsection{Bloch tight-binding model with further than nearest-neighbor hopping}

\label{ap:next_nearest}

At nearest and next-nearest neighbor range, the displacement vector
of Eq.~(\ref{eq:Ham_kernel}) runs over
\begin{equation}
\bm{\delta}=\pm\bm{a}_{1},\pm\bm{a}_{2},\pm\bm{a}_{3},\hspace{0.5cm}\pm(\bm{a}_{1}+\bm{a}_{2}),\ \pm(\bm{a}_{2}+\bm{a}_{3}),\ \pm(\bm{a}_{3}-\bm{a}_{1}).    
\end{equation}
Discrete rotation invariance then implies that
\begin{equation}
H_{\bm{k}jj'}=\delta_{jj'}\varepsilon_{j}+s_{\bm{k}jj'}t_{jj'}+s_{\bm{k}jj'}^{\mathrm{n}}t_{jj'}^{\mathrm{n}}
\end{equation}
where the next nearest-neighbor terms are defined by
\begin{equation}
t_{jj'}^{\mathrm{n}}\equiv e^{i\frac{\pi}{6}(j-j')}\,t_{(\bm{a}_{1}+\bm{a}_{2})jj'}
\end{equation}
\begin{equation}\begin{aligned}
s_{\bm{k}jj'}^{\mathrm{n}} & \equiv  e^{-i\frac{\pi}{6}(j-j')}\,\left(e^{-i\bm{k}\cdot(\bm{a}_{1}+\bm{a}_{2})}+e^{-i\bm{k}\cdot(\bm{a}_{2}+\bm{a}_{3})}e^{-i\frac{\pi}{3}(j\!-\!j')}+e^{-i\bm{k}\cdot(\bm{a}_{3}-\bm{a}_{1})}e^{-i\frac{2\pi}{3}(j\!-\!j')}+\right.\\
&  \quad\phantom{e-i\pi(j-j)}\left.e^{i\bm{k}\cdot(\bm{a}_{1}+\bm{a}_{2})}e^{-i\pi(j\!-\!j')}+e^{i\bm{k}\cdot(\bm{a}_{2}+\bm{a}_{3})}e^{-i\frac{4\pi}{3}(j\!-\!j')}+e^{i\bm{k}\cdot(\bm{a}_{3}-\bm{a}_{1})}e^{-i\frac{5\pi}{3}(j\!-\!j')}\right).
\end{aligned}\end{equation}
By direct substitution, we can verify the following properties:
\begin{equation}\begin{aligned}
s_{\bm{k}jj'}^{\mathrm{n}} & =  s_{\bm{k}-\!j'-\!j}^{\mathrm{n}}\\
s_{\bm{k}jj'}^{\mathrm{n}} & =  s_{M_{y}\bm{k}-j\!-\!j'}^{\mathrm{n}}\\
s_{\bm{k}j'j}^{\mathrm{n}} & =  \left(s_{\bm{k}jj'}^{\mathrm{n}}\right)^{*}e^{-i\pi(j\!-\!j')}
\end{aligned}\end{equation}
Imposing discrete rotation symmetry, particle-hole symmetry, and chiral
symmetry yields
\begin{equation}\begin{aligned}
t_{jj'}^{\mathrm{n}} & = \left(t_{jj'}^{\mathrm{n}}\right)^{*}\\
t_{jj'}^{\mathrm{n}} & = (-\!1)^{(j-j')}\ t_{j'j}^{\mathrm{n}}\\
t_{jj'}^{\mathrm{n}} & = (-\!1)^{(j-j'+1)}\ t_{-\!j'\,-\!j}^{\mathrm{n}}
\end{aligned}\end{equation}
It is straightforward to extend this to next-next-nearest neighbor
hopping and beyond.

\subsection{Phase boundaries of the two-band nearest-neighbor toy model}
Phase boundaries, which separate the Chern sectors, are lines in the phase diagram where the energy gap between the two bands vanishes for some momentum. This turns out to occur at $\Gamma$ and other high-symmetry momenta. We can determine the phase boundaries analytically and map out the complete phase diagram of the two-dimensional parameter space. We note that $E_{\bm{k}\pm}$, given in Eq.~(\ref{eq:two-band-energies}) of the main text, vanishes when both $\varepsilon+P_{\bm{k}}\tau=0$ and $Q_{\bm{k}}\tilde\tau=0,$ which are the terms under the square root. $Q_{\bm{k}}$ vanishes at the following momenta:
\begin{align}
(\bm{k}\cdot \bm{a}_1, \bm{k}\cdot \bm{a}_2) \in \big\{ (0,0), (0,\pi), (\pi, 0), (\pi, \pi), (2\pi/3, -2\pi/3), (-2\pi/3, 2\pi/3) \big\}.
\end{align}
The corresponding values of $P_{\bm{k}}$ at these points are $P_{\bm{k}} = 6,-2,-2,-2,-3,-3$. Therefore, when $\tilde{\tau}/\varepsilon \ne 0$, the phase boundaries occur along the lines $\tau/\varepsilon = -1/6, 1/3, 1/2.$
Attention should be paid to the special case where $\tilde{\tau}/\varepsilon = 0$.  In this case a gap closure occurs whenever $1 + \tau P_{\bm{k}}/\varepsilon = 0$ for some value of $\bm{k}$. Since $P_{\bm{k}}$ is a continuous function with a maximum value of $6$ and a minimum value of $-3$, this happens for some value of $\bm{k}$ whenever
\begin{eqnarray}
\tau/\varepsilon \ge 1/3 \mathrm{\phantom{and} or \phantom{and}} \tau/\varepsilon \le -1/6.
\end{eqnarray}
The analysis of these two cases gives us a complete picture of the phase boundaries. The Chern numbers themselves are given by the number of times the unit vector $\hat{h}_{\bm{k}}$ covers the unit sphere as the momentum is scanned through the BZ, namely
\begin{equation}
\mathcal{C}_{\pm} = \pm\frac{1}{2\pi}\int_{\text{BZ}}\! d^2\bm{k}\ \frac{1}{2}\hat h_{\bm{k}} \cdot\left(\partial_{k_x}\hat h_{\bm{k}} \times \partial _{k_y} \hat h_{\bm{k}}\right).
\end{equation}
The results are summarized in Fig.~\ref{fig:2bandPD}.


\end{document}